\documentclass[pra,twocolumn,preprintnumbers,amsmath,amssymb]{revtex4}

\usepackage{amscd}

\usepackage{graphicx}
\usepackage{dcolumn}
\usepackage{bm}

\newcommand{\sH}{\operatorname{H}} 
\newcommand{\metricd}{\operatorname{d}} 
\newcommand{\vect}[1] {\ensuremath{\vec{#1}}} 

\newcommand{\set}[1] {\ensuremath{\mathcal{#1}}} 

\newcommand{\state} [1] {\ensuremath{\mathbf{#1}}} 
\newcommand{\mment}[1] {\ensuremath{\mathbf{#1}}} 
\newcommand{\interaction}[1] {\ensuremath{\mathbf{#1}}} 
\newcommand{\map}[1]{\ensuremath{\mathcal{#1}}}

\newcommand{\spc}[1]{\ensuremath{\set{#1}}} 

\newcommand{\numberfield}[1]{\ensuremath{\mathbb{#1}}} 

\newcommand{\lamvec}{\ensuremath{\lambda_1, \lambda_2, \dots, \lambda_{M-1}}}

\newcommand{\veclam}{\ensuremath{\underline{\lambda}}} 

\newcommand{\cvect}[1] {\textsf{#1}} 

\newcommand{\lvect}[1] {\ensuremath{\mathbf{#1}}} 

\newcommand{\cmatrix}[1]{\textsf{#1}} 

\newcommand{\rmatrix}[1]{\ensuremath{#1}} 

\newcommand{\Bpi}{\bm{\pi}} 

\newcommand{\model}[1]{\ensuremath{\mathbf{#1}}} 

\newcommand{\orthant}{\ensuremath{\mathit{S}^{M-1}_+}} 
\newcommand{\hypersphere}{\ensuremath{\mathit{S}^{M-1}}} 

\addtocounter{secnumdepth}{-1} 

\newcommand{\var}{\ensuremath{\chi}} 
\newcommand{\vard}{\ensuremath{\xi}} 

\begin{document}



\title{An Information-Theoretic Approach to Quantum Theory,~I: \\
        The Abstract Quantum Formalism}

\author{Philip Goyal}
    \email{pg247@cam.ac.uk}
    \affiliation{ Astrophysics Group \\
Cavendish Laboratory \\
University of Cambridge }


\begin{abstract}
In this paper and a companion paper, we attempt to systematically
investigate the possibility that the concept of information may
enable a derivation of the quantum formalism from a set of
physically comprehensible postulates.  To do so, we formulate an
abstract experimental set-up and a set of assumptions based on
generalizations of experimental facts that can be reasonably taken
to be representative of quantum phenomena, and on theoretical ideas
and principles, and show that it is possible to deduce the quantum
formalism. In particular, we show that it is possible to derive the
abstract quantum formalism for finite-dimensional quantum systems
and the formal relations, such as the canonical commutation
relationships and Dirac's Poisson Bracket rule, that are needed to
apply the abstract formalism to particular systems of interest. The
concept of information, via an information-theoretic invariance
principle, plays a key role in the derivation, and gives rise to
some of the central structural features of the quantum formalism.

\end{abstract}

\maketitle

\section{Introduction}

Over the last two decades, a number of authors have expressed the
view that our efforts to develop an understanding of quantum theory
are impeded by a lack of understanding of the physical origin of the
quantum formalism, and that our efforts would thereby be
significantly aided by a systematic derivation of the formalism from
a set of physically comprehensible assumptions~\cite{Rovelli96,
Popescu-Rohrlich97, Fuchs02}. Furthermore, several authors have
proposed that the concept of information may be the key, hitherto
missing, ingredient which, if appropriately applied and formalized,
might make such a derivation possible~\cite{Wheeler89, Rovelli96,
Summhammer99, Zeilinger99, Fuchs02}.

The proposal that information might enable a derivation of quantum
formalism rests, to a significant degree, upon the recognition that
the concept of information plays a new and fundamental role in
quantum physics.  One way to see this is as follows.  In classical
physics, an experimenter presented with a system in an unknown state
can, in principle, perform an ideal measurement upon the system
which gives perfect knowledge about the state of the system.  Hence,
there is no fundamental distinction between the state and an ideal
experimenter's \emph{knowledge} of the state.  In quantum physics,
however, an ideal measurement~(or even a finite number of such
measurements performed upon an ensemble of identically-prepared
systems) provides only partial knowledge about the unknown state of
a quantum system. Hence, in sharp contrast to the situation in
classical physics, a fundamental distinction is drawn between the
state and the knowledge that the experimenter can conceivably have
of it. The concept of information then immediately assumes a
fundamental role through the natural attempt to quantitatively
\emph{relate} the two:~`How much information has been obtained by
the experimenter about the state?'

One of the earliest attempts to explore the role of information is
due to Wootters~\cite{Wootters80}. Suppose that Alice has a
Stern-Gerlach apparatus oriented at angle~$(\theta, \phi)$, and
attempts to communicate the angle~$\theta$ to Bob using spin-$1/2$
particles as follows.  Alice prepares~$n$ spin-$1/2$ particles in
the state~$|+\rangle_{\theta, \phi}$ using her Stern-Gerlach
apparatus, and sends the particles to Bob, who measures them using a
vertically-aligned Stern-Gerlach apparatus. The data he obtains
provides information about the outcome probabilities,~$P_1, P_2$, of
the measurement, where~$P_1$ is the probability of a spin emerging
in the positive channel. Since, from quantum theory,~$P_1 =
\cos^2(\theta/2)$, Bob thereby gains information about~$\theta$.
However, we can now ask the question:~suppose we did not know
quantum theory, and instead simply regard the experimental
arrangement as a way for Bob to learn about~$\theta$ by observing
the frequencies of the two possible outcomes of his Stern-Gerlach
apparatus; what function~$P_1(\theta)$ would maximize the amount of
information obtained by Bob about~$\theta$ for given~$n$?  Wootters
finds that, if the information is quantified using the Shannon
information measure, then, in the limit as~$n \rightarrow \infty$,
the function is~$P_1(\theta) = \cos^2(m\theta/2)$, where~$m\in
\numberfield{Z}^+$, a generalized form of Malus' law, which includes
the correct result as a special case.

Wootters' result is remarkable since it shows that, using the
standard inferential methods of probability theory and the
well-established Shannon information measure, and taking an
operational approach that assumes the probabilistic nature of
measurement outcomes, it is possible to make a correct, non-trivial
physical prediction concerning a quantum experiment from a plausible
information-theoretic principle. However, Wootters' attempt to
generalize this result in the direction of the quantum formalism
meets with limited success.

More recently, other attempts~\cite{Brukner99, Brukner02a,
Brukner02b, Summhammer99} have been made to examine and quantify the
gain of information in the measurement process, and which differ in
various ways from Wootters' approach, but which are also able to
derive the generalized form of Malus' law.  However, as with
Wootters' approach, they are unable to generalize their results to
obtain a significant part of the quantum formalism.

In contrast, several other recent approaches~\cite{Rovelli96,
Caticha98b, Caticha99b, Clifton-Bub-Halvorson03, Grinbaum03,
Grinbaum04} which involve the concept of information succeed in
deriving a significant fraction of the quantum formalism. However,
at the outset, these approaches make abstract assumptions of key
importance which are given no physical interpretation, and which
detract from the understanding of the physical origin of the quantum
formalism that can thereby be obtained. For instance, in the
approach described in~\cite{Caticha98b}, it is shown that, provided
one assumes that a complex number is associated with each
suitably-defined experimental set-up, Feynman's
rules~\cite{Feynman48} for combining complex probability amplitudes
can be derived from a set of plausible consistency conditions.
However, the choice of number field is not given a physical
interpretation, and an alternative choice of field, such as the
reals or quaternions, would lead to a different set of rules. In the
approaches described in~\cite{Clifton-Bub-Halvorson03, Grinbaum04},
a similar choice regarding the applicable number field is made at
the outset~\footnote{Specifically,
    (a)~in~\cite{Clifton-Bub-Halvorson03}, it is assumed that a physical
    theory can be accommodated within a~$C^*$-algebraic framework,
    which employs the complex number field, and~(b)~Grinbaum's
    Axiom VII~\cite{Grinbaum04} makes specific assumptions regarding
    the applicable number fields.
}.

In this paper and a companion paper~\cite{Goyal-QT2}~(hereafter
referred to as Paper~II), we attempt to build upon the insights
provided by Wootters' approach, and formulate an
information-theoretic principle and a set of physically
comprehensible assumptions from which it is possible to derive the
standard formalism of quantum theory. In particular, we obtain the
finite-dimensional abstract quantum formalism, namely~(a)~the von
Neumann postulates for finite-dimensional systems, (b)~the tensor
product rule for expressing the state of a composite system in terms
of the states of its sub-systems, and (c)~the result due to Wigner
that any symmetry transformation of a quantum system can be
represented by a unitary or antiunitary
transformation~\cite{Wigner-group-theory}.  In addition, we obtain
the formal rules of quantum theory~\footnote{The formal rules of
quantum theory can be categorized as follows:~(i)~\emph{Operator
Rules:}~the rules for writing down operators representing
measurements that, from a classical viewpoint, are measurements of
functions of other observables, (ii)~\emph{Commutation
Relations:}~the commutation relationships for measurement operators,
for example those operators representing measurements of position,
momentum, and components of angular momentum,
(iii)~\emph{Transformation Operators:}~explicit forms of the
operators that represent symmetry transformations~(such as
displacement) of a frame of reference, and
(iv)~\emph{Measurement--Transformation Relations:}~the relations
between measurement operators and the operators representing passive
transformations between physically equivalent reference frames.},
such as the canonical commutation relations, which are necessary to
apply the abstract formalism to obtain concrete models of particular
experimental set-ups.  We proceed as follows.

First, in Sec.~\ref{sec:F-idealised-set-up}, we describe an
idealized, abstract experimental set-up, which provides a general
framework within which particular experimental set-ups can be
described.  The preparations, interactions, and measurements that
are permitted in a given set-up are defined in an operational
manner.  This makes it possible to operationally specify set-ups,
where, like those set-ups ordinarily considered in quantum theory,
the preparation provides the maximum possible control over the
system insofar as predictions about the outcome probabilities of the
measurement are concerned, and the interactions only affect the
degrees of freedom of the state of the system that are under control
of the preparation.

Second, in Sec.~\ref{sec:F-statement-of-postulates}, we present a
set of postulates which concern the behavior of measurements
performed on the system, and which determine the theoretical
representation of measurements, the state of the system, and
physical transformations of the system.  The postulates are
formulated so as to be physically comprehensible, and an analysis of
their comprehensibility is presented in Sec.~\ref{sec:overview}. 
The key postulate is the \emph{Principle of Information Gain}, which
expresses the idea that, although different measurements yield
different information about the state of a system, they nonetheless
provide the same \emph{amount} of information about the state. That
is, although different measurements provide different perspectives
on a system, none is informationally privileged with respect to any
other.

Third, in Sec.~\ref{sec:D}, we show that, within the framework
provided by the abstract set-up, these postulates are sufficient to
derive the finite-dimensional abstract quantum formalism, apart from
the form of the temporal evolution operator.  In Paper~II, we
formulate an additional principle, the \emph{Average-Value
Correspondence Principle}, with which we obtain the form of the
temporal evolution operator and the formal rules of quantum theory.

In the course of the derivation, we find that the concept of
information, via the principle of information gain, gives rise to a
number of the key features of the quantum formalism, such as the
importance of square-roots of probability~(real amplitudes) and the
sinusoidal variation of probability with parameters, and plays a key
role in the restriction of possible transformations of state space
to unitary and antiunitary transformations.

We conclude in Sec.~\ref{sec:discussion} with a discussion of the
results.

\section{Experimental Set-up and Postulates}
\label{sec:F}

In this section, we shall first present an idealized, abstract
experimental set-up, which provides a general framework within which
particular experimental set-ups can be described.  We shall then
state a set of postulates which determines the abstract theoretical
model of the abstract experimental set-up.

\subsection{Abstract Experimental Set-up}
\label{sec:F-idealised-set-up}

\subsubsection{Introduction}

The description of an experimental set-up in a manner sufficiently
precise to enable modeling using the quantum formalism involves the
use of terms that are particular to the abstract language of the
quantum formalism.  For example, one speaks of a set-up that
prepares a system in a \emph{pure state}, but the concept of a pure
state has a specialized meaning which presupposes the quantum
formalism. However, since our goal is to derive the formalism, our
first task is to devise a way of defining, with sufficient
precision, what constitutes an experimental set-up without making
reference to such terms.

At the outset, we shall adopt, as background assumptions, the
following idealizations drawn from classical physics:
\begin{itemize}
\item[(a)]\emph{Partitioning}. The universe is partitioned into a
system, the background environment~(or simply,
the~\emph{background})~\footnote%
{The background environment of a systems is, by definition, that
part of the environment of a system which non-trivially influences
the behavior of the system, but which is not reciprocally affected
by the system.  For example, if a planet in the gravitational
field of a star is modeled as a test particle in a fixed
gravitational field of the star, then the planet~(test particle)
is the system, and the gravitational field is its background.  If
a part of the environment is reciprocally affected by the system,
the system is enlarged to include this part of the environment.
For example, if the reciprocal affect of the planet on the star is
relevant, the system is enlarged to include the star, and the star
and planet are regarded as interacting sub-systems within the
enlarged system.}
of the system, measuring apparatuses, and the rest of the
universe.
\item[(b)]\emph{Time}.  In a given frame of reference, one can
speak of a physical time which is common to the system and its
background, and which is represented by a real-valued
parameter,~$t$.
\item[(c)]\emph{States.} At any time, the system is in a definite
physical state, whose mathematical description is called the
mathematical state, or simply the \emph{state}, of the system. The
state space of the system is the set of all possible states of the
system.
\end{itemize}

\begin{figure*}
\begin{centering}
\includegraphics[width=6.75in]{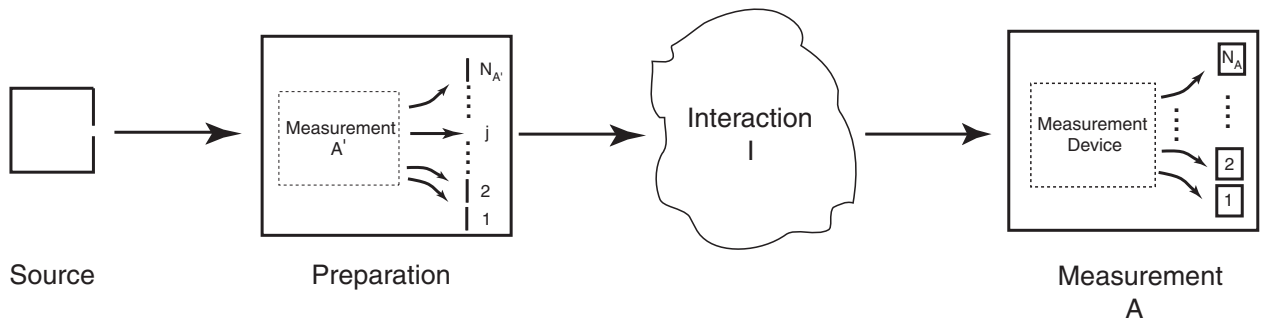}
\caption{\label{fig:idealised-set-up} An abstract, idealized
experimental-set up.  A physical system~(such as a silver atom) is
emitted from a source, passes a preparation step, and is then
subject to a measurement.   The preparation is implemented as a
measurement,~$\mment{A}'$, which has~$N_{A'}$ possible outcomes,
followed by the selection of those systems which yield some
outcome~$j$~($j=1,2, \dots, N_{A'}$). The measurement,~$\mment{A}$,
has~$N_{A}$ possible outcomes.  The measurement detectors are
assumed not to absorb the systems that they detect.  An
interaction,~$\interaction{I}$, may occur as indicated between the
preparation and measurement.}
\end{centering}
\end{figure*}

The general abstract experimental set-up that we shall consider is
shown in~Fig.~\ref{fig:idealised-set-up}. A \emph{source} provides
identical copies of a physical system of interest. A
\emph{preparation step} either selects or rejects the incoming
system. In a particular run of the experiment, a physical system
from the source passes the preparation, and is then subject to a
\emph{measurement} or \emph{measurements}.  In addition, following
the preparation, the system may undergo an \emph{interaction} with a
physical apparatus.

We shall only consider set-ups which satisfy particular
idealizations. In particular, we shall restrict consideration to
measurements that have the following properties:
\begin{itemize}
\item[(i)]\emph{Finiteness}:~the measurements yield a finite number of
possible outcomes,

\item[(ii)]\emph{Distinctness}:~the possible
outcomes of a measurement have distinct values,

\item[(iii)]\emph{Repetition Consistency}:~when a measurement is
immediately repeated, the same outcome is observed with certainty,
and

\item[(iv)]\emph{Classicality}:~~the measurements do not involve
auxiliary quantum systems.

\end{itemize}
In addition, we shall assume that interactions have the following
properties:
\begin{itemize}
\item[(i)]\emph{Identity-preserving}:~the interactions preserve the
identity of the system, and

\item[(ii)]\emph{Reversible and deterministic}:~the interactions are
reversible and deterministic at the level of the state of the
system, and so can be represented as one-to-one maps over state
space.

\end{itemize}

We shall also assume that the background of the system can be
adequately modeled within the classical framework insofar as its
internal dynamics is concerned.  For example, in the case of a
system in a background electromagnetic field, the field is assumed
to be modeled classically.  Similarly, we shall assume that
parameters which determine the measurement being performed~(the
orientation of a Stern-Gerlach apparatus, for instance) are
described classically as real-valued numbers.  In short, it is
assumed that the non-classicality is entirely concentrated in the
system and in its interactions with the background and the
measurement devices.

\subsubsection{Completeness of a Preparation}

The essential purpose of the experimental set-up illustrated in
Fig.~1 is to allow some property of a physical system to be studied
under controlled conditions~\footnote{The use of the word `property'
should be understood loosely here:~for example, one can, in both
classical and quantum physics, speak of the spatial and spin
properties of a particle with spin.}. Ideally, one would like to
prepare the system such that, immediately following the preparation,
one has as much knowledge as possible about the degrees of freedom
of the state of the system that are relevant to the property under
study, and one would like to interact with the system so that only
these degrees of freedom are affected. For example, if one wishes to
study the spin properties of a system, one would prepare the system
so that its spin direction is fixed~(in classical physics), or its
state is pure~(in quantum physics). Similarly, one would allow
uniform $\vect{B}$-field interactions since these only affect the
spin degrees of freedom of the system, but non-uniform
$\vect{B}$-field interactions would be excluded since they couple
spin and spatial degrees of freedom, and since spatial degrees of
freedom are not under control of the preparation.

Now, ordinarily, we rely upon a particular physical theory to tell
us which preparations are \emph{maximal} with respect to a given
measurement in the sense that they provide us with as much control
as physically possible over the degrees of freedom of the state of
the system that are relevant to predictions concerning the outcomes
of the given measurement, and which interactions are
\emph{compatible} with the preparation and measurement in the sense
of only affecting the degrees of freedom that are under control of
the preparation. However, since our goal is to derive the abstract
quantum formalism, where measurements and interactions are treated
purely in the abstract, it is necessary to find a way to establish
\emph{when} a preparation is maximal with respect to a given
measurement, and when an interaction is compatible with a
preparation and measurement, in a correspondingly abstract manner.

To do so, we make use of the fact that, in both classical and
quantum physics, a preparation is maximal with respect to a given
measurement if and only if the preparation is \emph{complete} in
that it renders the history of the system prior to the preparation
irrelevant insofar as predictions concerning the measurement
outcomes are concerned.  For example, in classical physics, if a
preparation places a system in a precisely known state~(which is, in
principle, possible), one has maximal degree of control over the
state, and the results of subsequent measurements performed on the
system are independent of the history of the system prior to the
preparation, so that the preparation is also complete. The converse
is also true.

In quantum physics, one encounters a similar situation. For example,
consider an experimental set-up where, in each run, a spin-1/2
system undergoes a preparation by a Stern-Gerlach measurement
device, and subsequently undergoes a Stern-Gerlach measurement. From
quantum theory, we know that the preparation in this case is maximal
with respect to the subsequent Stern-Gerlach measurement, and we
also know that the outcome probabilities of the measurement are
independent of the pre-preparation history of the spin-1/2 system,
so that the preparation is also complete.  The converse is also
true. More generally, if the preparation of a quantum system is
maximal with respect to a given projective measurement, then we know
from quantum theory that a system is prepared in a pure state, so
that the preparation is also complete with respect to the
measurement; and conversely.

Now, most importantly, unlike the notion of maximality, it is
straightforward to operationalize the notion of
completeness:~continuing with the example of the spin-$1/2$
experiment, if one models the data obtained from the measurement
in~$n$ runs of the experiment using a probabilistic
source~\footnote{A probabilistic source is a black box which, upon
each interrogation yields one of a given number of outcomes with a
given probability.}, one finds that, in the limit of large~$n$, the
outcome probabilities of the source are independent of
arbitrary pre-preparation interactions~%
\footnote{Here and subsequently, it is assumed that all interactions
with the system preserve the identity of the system.}
with the system.

Using this operationally-defined notion of completeness as a
basis, we shall see below that it is possible to give precise
expression to the idea that, roughly speaking, a pair of
measurements are examining the same property of the system from
different perspectives, and that an interaction is only
manipulating this particular property of the system.

\subsubsection{Definitions}

The measurements employed in the abstract set-up are chosen from a
\emph{measurement set},~$\set{A}$. As mentioned previously, it
will be assumed that each measurement has the property of
finiteness, which we shall now operationalize by saying that, when
the measurement is carried out on a system which has been emitted
from the source and has undergone arbitrary interactions
thereafter, the measurement generates one of a finite number of
\emph{possible outcomes}, a possible outcome being defined as one
that has a non-zero probability of occurrence.    It will also be
assumed that the measurement detectors do not absorb the systems
that they detect.

A preparation consists of a measurement that determines to which
outcome the incoming system belongs, followed by the selection of
the system if the measurement registers a given outcome, and the
rejection of the system otherwise.  If detectors that do not
absorb the detected systems are unavailable, a preparation can
instead be implemented using a measurement where one of the
detectors is removed.

Consider now an experiment~(Fig.~\ref{fig:idealised-set-up}) in
which a system from a source is subject to a preparation consisting
of measurement,~$\mment{A'}$, with~$N_{A'}$ possible outcomes, with
outcome~$j$ selected~($j=1,\dots, N_{A'}$), followed by
measurement~$\mment{A}$~(with~$N_A$ possible outcomes), without an
interaction in the intervening time.

Suppose that the data obtained in~$n$ runs of the experiment are
modeled by a probabilistic source with~$N_A$ possible outcomes,
whose most likely probabilities~(calculated on the basis of the
data) are given by~$\vect{P} = (P_1, P_2, \ldots, P_{N_A})$,
where~$P_i$ is the probability of the~$i$th outcome~$(i=1, 2, \dots,
N_A)$~
\footnote{As will be shown in Sec.~\ref{sec:D1-info-gain}, the
modeling process can be formalized using standard methods of
Bayesian data analysis. See~\cite{Sivia96}, for example, for a
general discussion on the subject.}.
If, for all~$j$, $\vect{P}$ is
 independent of arbitrary pre-preparation interactions
 with the system in the limit of large~$n$, the
preparation will be said to be \emph{complete} with respect to
 measurement~$\mment{A}$.  If the completeness condition also holds
true when~$\mment{A}$ and~$\mment{A}'$ are interchanged,
then~$\mment{A}$ and~$\mment{A}'$ will be said to form a
\emph{measurement pair}.

The set of measurements \emph{generated} by~\mment{A} forms a
measurement set,~$\set{A}$, which is defined as the set of all
measurements that~(i) form a measurement pair with~$\mment{A}$ and
that~(ii) are not a composite of other measurements in~$\set{A}$. An
important corollary of this definition is that two measurement sets
are either identical or disjoint.

Interactions that occur after the preparation step are chosen from
an \emph{interaction set},~$\set{I}$, which is defined as follows.
Suppose that, in the experiment of Fig.~\ref{fig:idealised-set-up},
an interaction,~$\interaction{I}$, occurs between the preparation
and measurement. If, for all~$\mment{A},\mment{A}' \in \set{A}$, the
preparation remains complete with respect to the subsequent
measurement, then~$\interaction{I}$ will be said to be
\emph{compatible} with~$\set{A}$ and the source. The set~$\set{I}$
is then defined as the set of all such compatible interactions.

If there are two experimental set-ups, each with a source
containing identical copies of the same physical system, with
respective disjoint measurement sets,~$\set{A}^{(1)}$
and~$\set{A}^{(2)}$, then the set-ups will be said to be
\emph{disjoint}. This makes precise the rough notion that the
set-ups examine different aspects of the same physical system.

\subsubsection{An example}

To illustrate the above definitions, consider again the spin-1/2
experiment, where silver atoms emerge from a source~(an
evaporator), pass through a Stern-Gerlach preparation device,
undergo an interaction, and finally undergo a Stern-Gerlach
measurement. In this case, the set,~$\set{A}$, generated by any
Stern-Gerlach measurement consists of all Stern-Gerlach
measurements of the form~$\mment{A}_{\theta, \phi}$,
where~$(\theta, \phi)$ is the orientation of the Stern-Gerlach
device.  However, measurements that are composed of two or more
Stern-Gerlach measurements are excluded from~$\set{A}$.

Consider now an interaction,~$\interaction{I}_{\theta_B, \phi_B,
t, \Delta t}$, consisting of a uniform $\vect{B}$-field acting
during the interval~$[t,t+\Delta t]$ in some direction~$(\theta_B,
\phi_B)$.  If such an interaction occurs between the preparation
and measurement, one finds that the completeness of the
preparation with respect to the measurement is preserved; that is,
the interaction is compatible with~$\set{A}$ and the system.
Hence, all interactions in which a uniform magnetic field acts
between the preparation and measurement are in the interaction
set,~$\set{I}$. However, interactions consisting of a non-uniform
$\vect{B}$-field do not preserve completeness~(viewed from the
quantum theoretic model, such interactions couple the spin and
position degrees of freedom of the system), and are therefore
excluded from~$\set{I}$.

Finally, to illustrate the concept of disjoint set-ups, consider a
source which emits a system consisting of two distinguishable
spin-1/2 particles on each run of an experiment, and consider two
set-ups where the first set-up has a measurement set~$\set{A}^{(1)}$
consisting of all possible Stern-Gerlach measurements performed on
one of the particles, and the second has a measurement
set~$\set{A}^{(2)}$ consisting of all possible Stern-Gerlach
measurements performed on the other particle. In this case, the two
measurement sets are disjoint. The set-ups themselves are
accordingly said to be disjoint, which precisely expresses the
notion that the two set-ups are examining distinct aspects of the
same physical system.

\subsection{Statement of the Postulates.}
\label{sec:F-statement-of-postulates}

Consider the idealized experiment illustrated in
Fig.~\ref{fig:idealised-set-up} in which a system passes a
preparation step that employs a measurement~$\mment{A}'$ in
measurement set~$\set{A}$, undergoes an
interaction,~$\interaction{I}$ in the interaction set~$\set{I}$, and
is then subject to a measurement,~$\mment{A}$, in~$\set{A}$. The
abstract theoretical model that describes this set-up satisfies the
following postulates.

\begin{enumerate}
\item[1.] \textbf{Measurements}
    \begin{itemize}
   \item[1.1]
        \emph{Finite and Probabilistic outcomes.}
            When any given measurement~$\mment{A} \in \set{A}$ is
            performed, one of~$N$~($N\geq 2$) possible outcomes are
            observed.
            The~$i$th outcome
            is obtained with probability~$P_i$~$(i=1,\dots,N)$,
            where~$P_i$ is determined by the preparation, interactions,
            and measurement.
    \item[1.2]
        \emph{Representation of Measurements.}
            For any given pair of measurements~$\mment{A},\mment{A}' \in \set{A}$, there
            exist interactions~$\interaction{I}, \interaction{I}' \in \set{I}$
            such that~$\mment{A}'$ can, insofar as probabilities of the
            outcomes
            and insofar as the output states of the measurement are concerned, be represented
            by an arrangement where~$\interaction{I}$ is immediately
            followed by~$\mment{A}$ which, in turn, is immediately
            followed by~$\interaction{I}'$.
    \end{itemize}
  \item[2.] \textbf{States}
    \begin{itemize}
    \item[2.1]
        \emph{States.}
    With respect to any given measurement~$\mment{A} \in \set{A}$, the
    state,~$\state{S}(t)$, of a quantum system at time~$t$ is
    given by~$(\vect{P}, \vect{\var})$, where~$\vect{P} = (P_1, P_2, \dots,
    P_N)$ and where~$\vect{\var} = (\var_1, \var_2, \dots,
    \var_N)$ is a set of~$N$ real degrees of freedom.

    \item[2.2]
        \emph{Physical interpretation of the~$\var_i$.}
    When measurement~$\mment{A} \in \set{A}$ is performed on a
    system in state~$\state{S}(t)$ and the outcome~$i$ is observed,
    there are additional outcomes that are objectively realized but
    unobserved:
        \begin{itemize}
        \item[(i)] one of two outcomes, labeled~$a$ and~$b$,
        which are obtained with respective
        probabilities~$P_{a|i} = Q^2_{a|i}$ and~$P_{b|i}= Q^2_{b|i}$,
        where~$Q_{a|i} = f(\var_i)$ and~$Q_{b|i}
        = \tilde{f}(\var_i)$,
        where~$f$ is not a constant function and~$f,\tilde{f}$
        have range~$[-1, 1]$, and
        \item[(ii)] one of two possible outcomes, with values
        labeled~$+$ and~$-$, which is determined by the sign
        of either~$Q_{a|i}$ or~$Q_{b|i}$ depending upon
        whether~$a$ or~$b$ has been realized.
        \end{itemize}

    \item[2.3]
        \emph{Information Gain.}
            When measurement~$\mment{A} \in \set{A}$ is performed on a
            system in any given unknown state~$\state{S}(t)$, the amount of
            Shannon-Jaynes information provided by the observed
            outcomes and the outcomes~$a$ and~$b$ about~$\state{S}(t)$ in $n$~runs of the experiment
            is independent of~$\state{S}(t)$ in the limit
            as~$n \rightarrow \infty$.

    \item[2.4]
        \emph{Prior probabilities.}
            The prior probability~$\Pr(\var_i|\text{I})$,
            where~I is the background knowledge of the experimenter prior to
            performing the experiment, is uniform for~$i=1, \dots, N$.
    \end{itemize}

\item[3.] \textbf{Transformations} Any transformation of a
prepared physical system, whether active~(due to temporal
evolution of the system), or passive~(a symmetry transformation
due to a change of the frame of reference), is represented by a
map,~$\map{M}$, over the state space,~$\spc{S}$, of the system.
    \begin{itemize}
    \item[3.1] \emph{One-to-one.} The map~$\map{M}$ is one-to-one.

    \item[3.2] \emph{Invariance.} The map~$\map{M}$ is such that,
    for any state~$\state{S} \in \spc{S}$, the observed outcome
    probabilities,~$P_1', P_2', \dots, P_N'$,
    of measurement~$\mment{A} \in \set{A}$ performed upon a system
    in state~$\state{S}'=\map{M}(\state{S})$ are unaffected if, in any
    representation,~$(\vect{P}, \vect{\var})=(P_i; \var_i)$, of the
    state~$\state{S}$ written down with respect to~$\mment{A}$,
    any arbitrary real constant,~$\var_0$, is added
    to each of the~$\var_i$.

    \item[3.3] \emph{Parameterized Transformations.} If a physical
    transformation is continuously dependent upon the real-valued
    parameter n-tuple~$\Bpi$, and is represented by the
    map~$\map{M}_{\Bpi}$, then~$\map{M}_{\Bpi}$ is continuously dependent
    upon~$\Bpi$.   If the physical transformation is a continuous
    transformation, then, for some value of~$\Bpi$,~$\map{M}_{\Bpi}$
    reduces to the identity.

    \item[3.4] \emph{Temporal Evolution.} The map,~$\map{M}_{t, \Delta t}$,
    which represents temporal evolution of a system in a time-independent
    background during the interval $[t, t+\Delta t]$, is such that any
     state,~$\state{S}$, represented as~$(P_i; \var_i)$, of
    definite energy~$E$,  whose observable degrees of freedom are
    time-independent, evolves to~$(P_i'; \var_i')$,
    where~$P_i'=P_i$ and~$\var_i' = \var_i - E\Delta t/\alpha$, where~$\alpha$
    is a non-zero constant with the dimensions of action.
    \end{itemize}
\item[4.]  \textbf{Consistency}  The posterior probability
distributions over~$\spc{S}$ that result from the following two
processes coincide in the limit as~$n \rightarrow \infty$:
    \begin{itemize}
    \item[(i)] inferring a posterior over~$\spc{S}$ based upon the
    objectively realized outcomes when the
    measurement~$\mment{A} \in \set{A}$
    is performed upon~$n$ copies of a system in state~$\state{S}$,
    and then transforming the posterior using~$\map{M}$, or
    \item[(ii)] inferring a posterior over~$\spc{S}$ based upon the
    objectively realized outcomes
    when the measurement~$\mment{A} \in \set{A}$
    is performed upon~$n$ copies of a system in state~$\map{M}(\state{S})$,
    \end{itemize}

\end{enumerate}

The above postulates, together with the Average-Value Correspondence
Principle~(AVCP), which will be given in Paper~II, suffice to
determine the form of the abstract quantum model for the abstract
set-up. From Postulates~1.1 and~1.3, it follows that, when any
measurement in~$\set{A}$ is performed on the system, one of~$N$
possible outcomes is observed. Accordingly, we shall denote the
abstract quantum model of such a set-up by~$\model{q}(N)$.

Finally, we shall need Postulates~5, below, in order to obtain a
rule, which we shall refer to as the \emph{composite systems
rule}, for relating the quantum model of a composite system to the
quantum models of its component systems:
\begin{enumerate}
\item[5.] \textbf{Composite Systems} Suppose that a system admits
a quantum model with respect to the measurement
set~$\set{A}^{(1)}$ whose measurements have~$N^{(1)}$ possible
observable outcomes, and admits a quantum model with respect to
measurement set~$\set{A}^{(2)}$ whose measurements have~$N^{(2)}$
possible observable outcomes, where the sets~$\set{A}^{(1)}$
and~$\set{A}^{(2)}$ are disjoint.

Consider the quantum model of the system with respect to the
measurement set~$\set{A} = \set{A}^{(1)} \times \set{A}^{(2)}$
that contains all possible composite measurements consisting of a
measurement from~$\set{A}^{(1)}$ and a measurement
from~$\set{A}^{(2)}$. If the states of the sub-systems are
represented as~$(P_i^{(1)}; \var_i^{(1)})$~$(i=1, 2, \dots,
N^{(1)})$ and~$(P_j^{(2)}; \var_j^{(2)})$~$(j=1, 2, \dots,
N^{(2)})$, respectively, then the state of the composite system
can be represented as~$(P_{ij}; \var_{ij})$,
where~$P_{ij}=P_i^{(1)}P_j^{(2)}$ and~$\var_{ij} = \var_i^{(1)} +
\var_j^{(2)}$.
\end{enumerate}

\section{Overview of the Postulates}
\label{sec:overview}

Many of the postulates described above can be seen to follow from
the quantum formalism, which provides some understanding of these
postulates. Accordingly, we shall first point out the relations
between these postulates and the quantum formalism.  We shall then
describe how the postulates can be physically understood.

\subsection{Postulates that follow from quantum theory}

Of the postulates enumerated above, all apart from
Postulates~2.2,~2.3,~2.4 and~4 can be seen to follow from the
quantum formalism.

Consider the quantum theoretical model of the abstract
experimental set-up.  Since the measurements in measurement
set~$\set{A}$ yield one of~$N$ possible distinct observable
outcomes, it follows that the state space of the quantum model is
$N$-dimensional. Furthermore, since a preparation~(implemented
using a measurement~$\mment{A}' \in \set{A}$) is complete with
respect to a measurement~$\mment{A} \in \set{A}$, it follows that
the system immediately following the preparation step is in a pure
state,~$\cvect{v} \in \numberfield{C}^N$.

According to the quantum formalism, measurement~$\mment{A}$ can be
represented by a Hermitian operator,~$\cmatrix{A}$.  With respect to
this measurement, the~$i$th component of~$\cvect{v}$ can be written
as~$P_i e^{i\phi_i}$, where~$P_i$ is the outcome probability of
outcome~$i$, so that the state can be represented as
\begin{equation} \label{eqn:cvectv-representation}
\cvect{v} = (P_1, \dots, P_N; \phi_1, \dots, \phi_N),
\end{equation}
or~$(P_i; \phi_i)$ for short, which yields Postulates~1.1 and~2.1.

In the quantum model, it is assumed that physical transformations
are represented by unitary or antiunitary transformations of state
space. Unitary and antiunitary transformations are one-to-one maps,
which gives Postulates~3 and~3.1.  To show Postulate~3.2, consider
the transformation of~$\cvect{v}e^{-i\phi_0}$ by the unitary
operator~$\cmatrix{U}$.  The transformed vector is
\begin{equation}
\cvect{v}' =e^{-i\phi_0}\cmatrix{U}\cvect{v}.
\end{equation}
However, the outcome probabilities of any measurement performed on
the system in state~$\cvect{v}'$ are independent of the overall
phase of~$\cvect{v}'$.  Therefore, these outcome probabilities are
unaffected if an arbitrary~$\phi_0 \in \numberfield{R}$ is added
to the~$\phi_i$, where~$\cvect{v}$ is represented as in
Eq.~\eqref{eqn:cvectv-representation}.

Postulate~3.3 is obtained in two parts.  First, if a physical
transformation depends continuously upon a set of real-valued
parameters, then it is represented by a unitary or antiunitary
transformation whose degrees of freedom also continuously depend
upon these parameters.  Second, continuous transformations are
represented by unitary transformations.  If a unitary
transformation is a continuous function of a set of real-valued
parameters, then it is possible that, for some values of these
parameters, the unitary transformation reduces to the identity.

From the unitary operator~$\cmatrix{U}_t(\Delta t) =
\exp(-i\cmatrix{H}_t \Delta t/\hbar)$ for the evolution of a system
during the interval~$[t, t+ \Delta t]$ in a time-independent
background, where~$\cmatrix{H}_t$ is the Hamiltonian operator at
time~$t$, it follows that a state~$\cvect{v}$ which is an eigenstate
of~$\cmatrix{H}_t$ evolves into
\begin{equation}
\cvect{v}' = e^{-iE \Delta t/\hbar} \cvect{v},
\end{equation}
where~$E$ is the energy of the state.  In the representation of
Eq.~\eqref{eqn:cvectv-representation}, the state~$(P_i; \phi_i)$
evolves to~$(P_i; \phi_i - E\Delta t/\hbar)$, and, since~$\cvect{v}$
and~$\cvect{v}'$ differ only by an overall phase, they are
observationally indistinguishable, which gives Postulate~3.4.

To show Postulate~1.2, suppose that one wishes to
represent~$\mment{A}'$ in terms of measurement~$\mment{A}$.
Consider an arrangement consisting of a unitary
transformation~$\cmatrix{U}$ immediately followed by
measurement~$\mment{A}$, followed immediately, in turn,
by~$\cmatrix{U}^\dagger$.  Suppose that measurements~$\mment{A}$
and~$\mment{A}'$ are represented by the operators~$\cmatrix{A}$
and~$\cmatrix{A}'$, respectively, where~$ \cmatrix{A}\cvect{v}_i =
a_i \cvect{v}_i$ and~$\cmatrix{A}'\cvect{v}'_i = a_i'
\cvect{v}'_i$.   Then, if we choose
\begin{equation}
\cmatrix{U} = \sum_i \cvect{v}_i \cvect{v}'^\dagger_i,
\end{equation}
this arrangement behaves precisely the same as
measurement~$\mment{A}'$ insofar as the probabilities of the
observed outcomes and insofar as the corresponding output states are
concerned.  To see this, note that, if the input state to the
arrangement is~$\sum_i c_i' \cvect{v}_i'$~(the~$c_i'$ being complex
constants, such that~$\sum_i |c_i'|^2 = 1$) the
state~$\cmatrix{U}\sum_i c_i'\cvect{v}_i' = \sum_i c_i'\cvect{v}_i$,
and therefore measurement~$\mment{A}$ yields outcome~$i$ with
probability~$|c_i'|^2$ and yields corresponding state~$\cvect{v}_i$
up to an irrelevant overall phase.  The final output state of the
arrangement is therefore~$\cmatrix{U}^\dagger \cvect{v}_i =
\cvect{v}_i'$. Hence, the arrangement behaves precisely as would
measurement~$\mment{A}'$ performed directly on a system in
state~$\cvect{v}'$ in respect of the probabilities of observed
outcomes~$1, 2, \dots, N$ and in respect of the output states.

Finally, by considering the tensor product~$\cvect{v} =
\cvect{v}^{(1)} \otimes \cvect{v}^{(2)}$ where~$\cvect{v}^{(1)}
\in \numberfield{C}^{N_1}$ and $\cvect{v}^{(2)} \in
\numberfield{C}^{N_2}$ are the states of two sub-systems,
and~$\cvect{v} \in \numberfield{C}^{N}$, with~$N=N_1N_2$, is the
state of the composite system, one finds that Postulate~5 follows
at once.

\subsection{Physical Comprehensibility of the postulates.}
\label{sec:physical-comprehensibility-of-the-postulates}

When formulating the postulates, our goal has been to maximize their
physical comprehensibility. For the purposes of discussion, it is
helpful to distinguish two \emph{levels} of physical
comprehensibility. First, at the minimum, a comprehensible postulate
is one that can be transparently understood as a simple assertion
about the physical world.  If this is the case, we shall say that
the postulate has the property of \emph{transparency}. Second, a
postulate has an additional level of comprehensibility if it can
also be traced to well-established experimental facts and physical
ideas or principles~(\emph{traceability}).

To illustrate these ideas, consider the example of Einstein's
postulate of the constancy of the speed of light.  The postulate can
be transparently understood as the simple assertion that
measurements of the speed of light in different inertial frames will
yield the same result.  In addition, the postulate can also be
understood as a direct generalization of the well-established
results of the Michelson-Morley experiment, the generalization being
achieved by an appeal to the general principle of the uniformity of
nature. Hence, the postulate is both transparent and traceable.

Since the assumptions underlying classical physics are transparent
and traceable to well-established experimental facts and theoretical
ideas, and since these assumptions remain fundamental to the way in
which we conceptualize the physical world, we attempt to preserve
them as far as possible in the face of quantum phenomena.
Accordingly, we draw the majority of the postulates from classical
physics, either by taking fundamental features of the theoretical
framework of classical physics and modifying these, if necessary, in
light of experimental facts that are characteristic of quantum
phenomena, or by transposing particular features of the classical
models of physical systems into the quantum realm via a
classical--quantum correspondence argument. Furthermore, in our
treatment of information, we use the standard inferential methods of
probability theory, and employ the conceptually and mathematically
well-established framework of Shannon information theory.   The
remaining assumptions, which have no obvious classical counterparts,
are based on experimental facts that are characteristic of quantum
phenomena but have no classical analog, or are based on novel
theoretical ideas and principles.

In our discussion below, we shall divide the postulates
into~(i)~postulates that are adopted from classical physics, or are
modified therefrom in light of experimental facts characteristic of
quantum phenomena, (ii)~postulates that are obtained through a
classical-quantum correspondence argument, and~(iii)~novel
postulates with no classical counterparts.

\subsubsection{Postulates adopted from classical physics.}

A classical model of a physical system is based upon the
\emph{partitioning}, \emph{time} and \emph{states} background
assumptions given earlier, and these are adopted unchanged in the
abstract quantum model.  The classical model additionally makes the
following additional key assumptions:
\begin{enumerate}
    \item[A] \emph{Measurements.}
        \begin{itemize}
        \item[A1] \emph{Operational Determinacy.} The outcome of a
        measurement performed on the system is determined by
        experimentally-controllable variables.
        \item[A2] \emph{Continuum.} The values of the possible outcomes
        of a measurement form a real-valued continuum.
        \end{itemize}

    \item[B] \emph{States.}

        \begin{itemize}
        \item[B1] \emph{Determinacy.} The state of the system and
        a theoretical description of a measurement that is performed on
        the system determine the measurement outcome.
        \end{itemize}

    \item[C] \emph{Transformations.}
        \begin{itemize}
        \item[C0] \emph{Mappings.} Physical transformations of the system,
         either due to temporal evolution or due to a passive change of
         frame of reference,  are represented by mappings over the space
          of states.
        \item[C1] \emph{One-to-one.} The mappings are one-to-one.
        \item[C2] \emph{Continuity.} If a map represents
        a physical transformation that depends continuously upon a
        real-valued set of parameters, then the map is continuously
        dependent upon these parameters.
        \item[C3] \emph{Continuous transformations.} If a map represents
        a continuous transformation~(such as temporal evolution) that
        depends continuously upon a set of real-valued parameters, then,
        for some value of these parameters, the map reduces to the
        identity.
        \end{itemize}
\end{enumerate}

We remark that the measurements mentioned in~A1--2 are idealized,
fundamental measurements, such as measurements of the position of
a particle, which, in the framework of classical physics, are
assumed to yield a continuum of possible outcomes~\footnote{One
can construct procedures which, for example, classify a particle
as being in one of a discrete~(finite or countably infinite)
number of regions of space, but, although one might describe such
a procedure as a `measurement', it is not regarded a fundamental
measurement in the classical framework.}. Similarly, although
fundamental measurements of a physical quantity in a particular
situation~(such as the frequency of a bound membrane) may take a
discrete number of possible values, it is assumed that the
discreteness arises through the particular boundary conditions
that are applicable, rather than being an intrinsic feature of the
measurements themselves.

We also remark that, in~C0-C3, it is assumed that physical
transformations of a physical system are deterministic and
reversible, which prevents the description of irreversible or
indeterministic transformations within the classical framework at
a fundamental level.

First, we consider those postulates which adopt classical
assumptions unchanged. Postulates~3 and~3.1 correspond,
respectively, to assumptions~C0 and~C1, while Postulate~3.3 is a
combination of assumptions~C2 and~C3.

Second, in light of the results of experiments involving quantum
systems~(such as Stern-Gerlach measurements on silver atoms), it is
reasonable to modify assumptions~A1,~A2 and~B1 as follows:
    \begin{itemize}
    \item[A1$'$] \emph{Probabilistic operational determinacy.} The
    data obtained when a measurement is performed on the system
    are best modeled by a probabilistic source whose outcome
    probabilities are determined by experimentally-controllable variables.
    \item[A2$'$] \emph{Finiteness.} A measurement performed on a system
    has a finite number of possible outcomes.
    \item[B1$'$] \emph{Probabilistic determinacy.} The state of the
    system and a theoretical description of a measurement that is
    performed on the system only \emph{probabilistically} determine
    the measurement outcome.
    \end{itemize}
We emphasize that, although these modifications are reasonable, they
are not the only possibilities consistent with the experimental
facts. For example, the probabilistic operational determinacy that
one finds empirically can be accommodated in at least two ways.
First, one can assume that the state of the system does, in fact,
determine the outcome of a measurement performed upon the system,
but that one cannot, for some reason, control all of the relevant
degrees of freedom of state.  Second, one can assume that the
degrees of freedom of the state only determine the probability that
a measurement yields a particular value.  In this instance, we have
taken the latter option.

These modified assumptions are contained within Postulates~1.1
and~2.1.  Specifically, Postulate~1.1 contains assumption~A1$'$
and~A2$'$, while Postulate~2.1 incorporates assumption~B1$'$.

\subsubsection{Postulates obtained through classical-quantum
correspondence.}

A general guiding principle in building up a quantum model of a
physical system is that, in an appropriate limit, the predictions of
the quantum model of the system stand in some one-to-one
correspondence with those of a classical model of the system.   By
establishing such a correspondence between the quantum and classical
models of a particle, we shall transpose several elementary
properties of the classical model across to the quantum model and
then, by generalization, to the abstract quantum
model,~$\model{q}(N)$.

Consider an experiment in which a position measurement is used to
prepare a particle at time~$t_0$, and a position measurement is
subsequently performed at time~$t_1$, during which interval a
potential~$V(\vect{r},t)$ is assumed to act.  When such an
experiment is actually performed, one necessarily uses position
measurements with a finite number of possible outcomes. In this
case, the experimental results~(where, for instance, an electron
passes through a sub-micron aperture, is subject to electric-field
interactions, and is subsequently detected on a screen) support
the conclusion that, if these coarse position measurements are of
sufficiently high spatial resolution, the preparation is, to a
very good approximation, complete with respect to the subsequent
measurement.

Suppose, then, that a coarse position measurement with~$N$
possible outcomes is used to implement both the preparation and
measurement steps, and further let us suppose that the coarse
measurement is such that the probability that a detection is
obtained in any run of the experiment is very close to unity.
Further, let us suppose that the coarse measurement is of
sufficient resolution that the preparation can be regarded as
being complete with respect to the measurement.  Then we can form
a quantum model, which we shall denote~$\model{q}^*(N)$, within
the framework of the abstract quantum model~$\model{q}(N)$, which
approximately describes the experiment after time~$t_0$.

By Postulate~1.1 and the assumption~B1$'$ above, the
state,~$\state{S}(t_1)$, of the system immediately prior to the
coarse position measurement determines the probability
n-tuple,~$\vect{P}(t_1)=(P_1, \dots, P_N)$, where~$P_i$ is the
probability of detection at the $i$th detector, which characterises
the data obtained from the coarse position measurement.

If the above experiment is repeated, except that the coarse
position measurement is delayed until time~$t_2$,
then~$\state{S}(t_1)$, together with a theoretical representation
of any interaction in the interval~$[t_1, t_2]$, must~(by
assumption~B1$'$) enable the prediction of the probability
n-tuple~$\vect{P}(t_2)$ that describes the coarse position
measurement data obtained at time~$t_2$.  To determine what
additional degrees of freedom the state~$\state{S}(t_1)$ must
contain in order to make this prediction possible, consider the
classical limit.

Suppose that~$m$ is increased towards values characteristic of
macroscopic bodies. Under the assumption made above, the
preparation is complete with respect to the measurement, so that
the system continues to be well-described by the
model~$\model{q}(N)$ even in this classical limit.
However, as~$m$ tends towards macroscopic values, it is reasonable
to expect that the system will increasingly behave in accordance
with its classical model between times~$t_1$ and~$t_2$. That is,
in this classical limit, we expect that~$\vect{P}(t_2)$, which is
determined in the quantum model in terms of~$\vect{P}(t_1)$ and
the other degrees of freedom in~$\state{S}(t_1)$, will coincide
with the n-tuple~$\vect{P}^{(\text{CM})}(t_2)$ that is predicted
by a classical model of a particle of mass~$m$ moving in the same
potential.

The relevant classical model in this situation is a particle
ensemble model.  For such an ensemble model, one can choose to
describe an ensemble for the case of given total energy by means
of a probability density function over phase space, and to
describe the evolution of this function using Newton's equations
of motion. Alternatively, one can employ the Hamilton-Jacobi
model, which is physically equivalent.  We choose the latter since
it is more easily described on a discrete spatial lattice.

In the Hamilton-Jacobi model, the state of the ensemble is given
by~$(P(\vect{r},t), S(\vect{r}, t))$, which satisfies the
Hamilton-Jacobi equations,
\begin{equation}\label{eqn:HJ}
\begin{gathered}
\frac{\partial P}{\partial t} + \nabla. \left(
                        \frac{1}{m} P \, \nabla  S
                                        \right) = 0 \\
\frac{1}{2m} \left( \nabla S \right)^2 + V(\vect{r}, t) = -
\frac{\partial S}{\partial t}.
\end{gathered}
\end{equation}
In the case of coarse position measurements with~$N$ possible
outcomes, we shall use the discretized form of the Hamilton-Jacobi
state,~$(\vect{P}^{(\text{CM})}; S_i)$, with~$i=1, \dots, N$, and
with~$\vect{P}^{(\text{CM})} = (P_1^{(\text{CM})}, \dots,
P_N^{(\text{CM})})$, where~$P_i^{(\text{CM})}$ is the probability
that the position measurement yields a detection at the~$i$th
measurement location, and~$S_i$ is the classical action at the~$i$th
measurement location.

In order that the predictions of the quantum and classical models
agree in the classical limit, the quantum
state~$\state{S}(t)$~($t>t_0$) must contain degrees of freedom
which encode~$N$ quantities, which we shall
denote~$S^{(\text{QM})}_1, \dots, S^{(\text{QM})}_N$, which, in
the classical limit, are equal to the~$S_i$.  Equivalently, we
shall assume that~$\state{S}$ contains~$N$ dimensionless real
quantities,~$\var_1, \dots, \var_N$, such that~$S^{(\text{QM})}_i
= \alpha\var_i$, where~$\alpha$ is a constant with dimensions of
action.

From the above discussion, in the model~$\model{q}^*(N)$, the
state,~$\state{S}$, is given by~$(\vect{P}, \vect{\var})$,
where~$\vect{\var}= (\var_1, \dots, \var_N)$.  Postulate~2.1
directly generalizes this statement to the abstract
model~$\model{q}(N)$.

We now observe that the Hamilton-Jacobi model has the following
properties, which can be readily verified from Eq.~\eqref{eqn:HJ}:
\begin{itemize}
\item[1.] \emph{Invariance.} The evolution of the
state~$(\vect{P}^{(\text{CM})}(t_1); S_i(t_1))$ to the
state~$(\vect{P}^{(\text{CM})}(t_2); S_i(t_2))$ is such
that~$\vect{P}^{(\text{CM})}(t_2)$ is unchanged if an arbitrary
real constant,~$S_0$, is added to each of the~$S_i(t_1)$.

\item[2.] \emph{Temporal Evolution.} In a time-independent
background, a state,~$(\vect{P}^{(\text{CM})}(t); S_i(t))$ whose
observable degrees of freedom are time-independent, evolves in
time~$\Delta t$ to the state~$(\vect{P}^{(\text{CM})}(t); S_i(t) -
E\Delta t)$, where~$E$ is the total energy of the system.

\item[3.] \emph{Composite Systems.} If, with respect to position
measurements along the~$x$ and~$y$ axes, the Hamilton-Jacobi state
of a particle is~$(P^{CM(x)}_i, S^{(x)}_i)$ and~$(P^{CM(y)}_j,
S^{(y)}_j)$, respectively, then, with respect to $xy$-position
measurements, its state is~$(P^{CM(xy)}_{ij}, S^{(xy)}_{ij}) =
(P^{CM(x)}_i P^{CM(y)}_j, S^{(x)}_i + S^{(y)}_j)$
\end{itemize}
Furthermore, from the first property, since the zero-value of
the~$S_i$ is conventional and therefore has no physical correlate,
the prior probability~$\Pr(S_i|\text{I})$ must be invariant under
arbitrary changes of the zero-value of the~$S_i$, where~$I$
represents the state of knowledge of the experimenter prior to
performing a measurement on the system.  The uniform prior is the
only prior that has this invariance property. Therefore, the
prior~$\Pr(S_i|\text{I})$ is uniform, which we shall list as a
fourth property:
\begin{itemize}
\item[4.] \emph{Prior Probabilities.} The
prior~$\Pr(S_i|\text{I})$ is uniform~($i=1, 2, \dots, N$),
where~$I$ represents the state of knowledge of the experimenter
prior to performing a measurement on the system.
\end{itemize}

On the assumption of the above correspondence between the
Hamilton-Jacobi model and the model~$\model{q}^*(N)$, it is now
possible to transpose these properties to the
model~$\model{q}^*(N)$ in the classical limit.

For example, Postulate~2.4 is obtained as follows.  First, by
using the relation~$\Pr(S_i|\text{I}) |dS_i| =
\Pr(\var_i|\text{I}) |d\var_i|$, it follows that
\begin{equation} \label{eqn:S-f-phi-relation}
\Pr(\var_i|\text{I}) \left| dS_i/d\var_i \right|^{-1} =
\Pr(S_i|\text{I}).
\end{equation}
Then, using the correspondence relation that~$S_i = \alpha\var_i$
in the classical limit, and noting that~$\Pr(S_i|\text{I})$ is
uniform~(property~4, above), we conclude that, in the classical
limit, the model~$\model{q}^*(N)$ satisfies the condition
that~$\Pr(\var_i|\text{I})$ is a constant. Second, the assumption
is made that this condition holds for the model~$\model{q}^*(N)$
not only in the classical limit but also for microscopic values
of~$m$ and, even more generally, that it holds for the abstract
quantum model~$\model{q}(N)$.

Postulates~3.2,~3.4 and~5 are obtained in a similar manner by
using the above correspondence,~$S_i = \alpha\var_i$, to transpose
the first three properties to the model~$\model{q}^*(N)$ in the
classical limit, and then making the assumption that the
transposed properties hold more generally for the abstract quantum
model~$\model{q}(N)$.

\subsubsection{Novel Postulates}

Below, we shall describe the four novel postulates, namely
Postulates~1.2,~2.2,~2.3 and~4.

\medskip
\paragraph{Postulate~1.2: Representation of Measurements.}

Consider an experiment in which Stern-Gerlach preparations and
measurements are performed upon silver atoms, and where the
set~$\set{A}$ consists of the elements~$\mment{A}_{\theta, \phi}$
representing Stern-Gerlach measurements in the direction~$(\theta,
\phi)$.  In this experiment, if an interaction consisting of a
uniform magnetic field acts between the preparation and
measurement, one finds that both the probabilities of the observed
outcomes are the same as would be obtained if a different
measurement had been done with the solenoid absent.

Using this observation, one finds that it is possible to implement
the measurement~$\mment{A}_{\theta, \phi}$ using any given
measurement~$\tilde{\mment{A}} \in \set{A}$ if followed
immediately before and after by suitable interactions.  The
implementation behaves precisely as~$\mment{A}_{\theta, \phi}$
insofar as the probabilities of observable outcomes~$1$ and~$2$,
and the corresponding output states, are concerned. Postulate~1.2
can be regarded as a plausible generalization of this observation.

\medskip
\paragraph{Postulate~2.2: Physical interpretation of the~$\var_i$.}

According to Postulate~2.1, the state~$S(t)$, written with respect
to some measurement~$\mment{A} \in \set{A}$, consists of the
pair~$(\vect{P}, \vect{\var})$, where~$\vect{P}$ contains the
probabilities of the observed outcomes, and~$\vect{\var}$ is an
ordered set of real-valued degrees of freedom.  Hence, the state
consists of a mixture of probabilities and degrees of freedom
unconnected to probabilities. Postulate~2.2 is motivated by the
aesthetical desideratum that a quantum state consist, as far as
possible, of probabilities of events, rather than being such a
mixture.

Accordingly, we postulate that~$\var_i$ encodes the probabilities of
some events, labeled~$a$ and~$b$.  Hence, when
measurement~$\mment{A}$ is performed on the system, one of
$2N$~possible outcomes is obtained, with probabilities determined by
the state of the system.  Since, by Postulate~1.1, the probabilities
of the observed outcomes of measurement~$\mment{A}$ are determined
by the~$P_i$, we are forced to postulate that, for some reason to be
investigated later, the outcomes~$a$ and~$b$ are not observed by the
experimenter.

Now, we make the reasonable assumption that the abstract quantum
framework being developed is capable of modeling the behavior of a
photon when subject to polarization measurements, and that this
model will agree with the predictions of electromagnetism under a
particle interpretation.  Now, an electromagnetic plane wave of
constant amplitude moving along the $+z$-direction is described by
the vector-valued function~$\vect{E} = E_0( \cos\theta\,\vect{i} +
\sin\theta\,\vect{j})$, and the information about the polarization
of the wave is contained in~$(\cos\theta, \sin\theta)$ with respect
to polarization measurements in the $xy$-plane. In the particle
interpretation, the probability that a photon will pass through a
polarizer whose axis points along the $x$-axis or~$y$-axis is given
by~$\cos^2\theta$ or~$\sin^2\theta$, respectively. The key feature
which we wish to abstract from this example is that, since the map
from~$(\cos\theta, \sin\theta)$~(the `state-level')
to~$(\cos^2\theta, \sin^2\theta)$~(the `probability-level') is
many-to-one, the computed probabilities are \emph{not} the
fundamental quantities when describing the state of the photon.
Rather, the more fundamental quantities are~$\cos\theta$
and~$\sin\theta$, which we can regard as square roots of probability
in the range~$[-1, 1]$, which are squared to obtain probabilities.

To incorporate this two-layered feature into the abstract quantum
model, we assume that, following the realization of outcome~$a$
or~$b$, one of two outcomes, labeled~$+$ and~$-$, is obtained. This
ensures that one binary-valued degree of freedom is associated with
each of the $2N$~possible probabilistically-determined outcomes.
Furthermore, we assume that the value of~$\var_i$ \emph{determines}
whether~$+$ or~$-$ is obtained via the sign of either~$Q_{a|i}$
or~$Q_{b|i}$, depending upon whether~$a$ or~$b$ was obtained,
where~$P_{a|i} = Q^2_{a|i}$ and~$P_{b|i} = Q^2_{b|i}$.  In summary,
the quantum state consists of the~$N$ probabilities~$P_1, \dots,
P_N$ and the~$2N$ quantities~$Q_{a|1}, Q_{b|1}, \dots, Q_{a|N},
Q_{b|N}$ which encode the probabilities~$P_{a|1}, P_{b|1}, \dots,
P_{a|N}, P_{b|N}$ and encode the values of the $2N$~binary-valued
degrees of freedom.

In Sec.~\ref{sec:discussion-of-postulate-2-2}, we sketch some
ideas which help to provide a better physical understanding of
this postulate.

\medskip
\paragraph{Postulate~2.3: Principle of Information Gain.}

Postulate~2.3 asserts that, in the arrangement of
Fig.~\ref{fig:idealised-set-up}, if measurement~$\mment{A} \in
\set{A}$ is performed on a system in any unknown
state~$\state{S}(t)$, then, in~$n$ runs of the experiment, the
amount of information provided by the probabilistically-determined
outcomes~(namely, one of~$1, \dots, N$, followed by either~$a$
or~$b$) about~$\state{S}(t)$ is independent of~$\state{S}(t)$ in
the limit as~$n \rightarrow \infty$. This postulate can be
understood physically as follows.

Suppose that, in trial~$1$ of $n$~runs of an experiment, a
measurement~$\mment{A}$ is performed on a system in
state~$\state{S}(t)$, and suppose that trial~$2$ is identical to
trial~$1$ except that measurement~$\mment{A}'$ is performed
instead of~$\mment{A}$. Now, by Postulate~1.2, trial~$2$ is
equivalent~(insofar as the probabilities of the
probabilistically-determined outcomes are concerned) to trial~$2'$
consisting of $n$~runs of an experiment where a system in
state~$\state{S}(t)$ is sent through an arrangement consisting of
a suitable physical interaction with the system, represented by
map~$\map{M}$~(Postulate~3), followed by measurement~$\mment{A}$,
followed by another physical interaction.

The data obtained in trials~$1$ and~$2$ provides information~(via
the Shannon-Jaynes entropy functional, as we shall later detail)
about~$S(t)$.  Furthermore, since the data obtained in trials~$2$
and~$2'$ is statistically identical~(as ensured by Postulate~1.2),
the amount of information obtained about~$S(t)$ in trial~$2$ is
asymptotically equal to the amount of information obtained
about~$S'(t)= \map{M}\left(S(t)\right)$ in trial~$2'$

Now, suppose that, in one of the two trials~$1$ and~$2$, the data
obtained yields more information about the state~$S(t)$ than in
the other trial.  This implies that, in the trials~$1$ and~$2$,
one of the two measurements~$\mment{A}$ and~$\mment{A}'$ is
privileged compared to the other insofar as the amount of
information that it yields about~$\state{S}(t)$. Although this
possibility cannot be ruled out \emph{a priori}, we make the
intuitively plausible assertion that, although these different
measurements provide different perspectives on the system, these
perspectives are not informationally privileged. Postulate~2.3
ensures that the amount of information obtained in trials~$1$
and~$2'$ is asymptotically equal and, therefore, that the amount
obtained in trials~$1$ and~$2$ is equal.  That is, Postulate~2.3
can be understood as arising from the requirement that no
measurement in the measurement set provides an informationally
privileged perspective on the system.

In order to quantify the amount of information gained, the
Shannon-Jaynes entropy functional~(also known as the relative
entropy) has been used~(see Eq.~\eqref{eqn:Shannon-Jaynes}), which
is the continuum generalization of the Shannon entropy~\footnote{%
The Shannon entropy,~$H(P_1, \dots, P_M) = - \sum_i P_i \ln P_i$
leads, via a straightforward continuum limit
argument~\cite{Jaynes63} to the Shannon-Jaynes entropy,~$H[p(x)]
=-\int p(x) \ln\left(p(x)/\mu(x)\right) dx$, of a probability
density function~$p(x)$, where~$\mu(x)$ is a measure over~$x$.  If
the Shannon-Jaynes entropy is used in the principle of maximum
entropy, then, in the absence of any data, the principle leads to
the assignment~$p(x) = \mu(x)$, which leads to the interpretation
that~$\mu(x)$ is the prior probability,~$\Pr(x|\text{I})$, where~I
symbolizes one's knowledge prior to obtaining the
data~(see~\cite{Probability-Theory-Jaynes}, \S~12.3).
The functional~$-\int p(x) \ln p(x) \,dx$ is often quoted as the
continuum generalization of the Shannon entropy, and indeed was
stated~(without proof) by Shannon in his foundational
paper~\cite{Shannon48}. However, a careful argument shows that the
correct continuum form is the Shannon-Jaynes entropy. The
Kullback-Leibler distance~(or the relative entropy) has the same
form as the Shannon-Jaynes entropy, but is generally not accompanied
by the interpretation of~$\mu(x)$ as the measure or prior
over~$x$.}.
Although other discrete information measures, such as the R\'enyi or
Tsallis entropies~\cite{Renyi65,Tsallis88}, have been proposed, the
Shannon-Jaynes entropy is preferred here since the Shannon entropy
has the clearest axiomatic basis~(being derivable from a set of
intuitively reasonable postulates~\cite{Shannon48, Khinchen57a,
Faddeev57}) and has strong indirect support through applications in
communication theory and through the many successes of the maximum
entropy method~(see~\cite{Jaynes57a, Jaynes57b}, for example), of
which it forms the basis.

In Sec.~\ref{sec:discussion-of-postulate-2-3}, we shall develop a
better understanding of this postulate and describe some of its
interesting consequences.

\medskip
\paragraph{Postulate~4: Consistency.}

A fundamental requirement of a theoretical model is that it be
internally consistent.  That is, if it is possible to make a
particular prediction via two distinct calculational pathways, the
predictions obtained must agree.

Postulate~4 considers the particular situation where one attempts to
calculate a posterior probability distribution over state space on
the basis of the objectively realized outcomes~(see Postulate~2.2)
in~$n$ runs of an experiment in which a measurement,~$\mment{A}$, is
performed on a system.

In particular, one can arrive at the posterior,~$p'(\state{S})$,
via two calculational pathways:
\begin{equation*}
\begin{CD}
 \state{S} @>\text{Map~\map{M}}>>\state{S}'=\map{M}(\state{S}) \\
 @V{\text{Measurement~\mment{A}}}VV                       @VV{\text{Measurement~\mment{A}}}V\\
 p(\state{S}) @>\text{Map~$\map{M}^*$}>> p'(\state{S})
\end{CD}
\end{equation*}
In the first route, in a given run of the experiment,
state~$\state{S}$ is first transformed to state~$\state{S}' =
\map{M}(\state{S})$, and then one performs measurement~$\mment{A}$
on the system.  On the basis of the data obtained in~$n$ runs, one
then calculates a posterior probability distribution over state
space. In the second route, in a given run, one first performs the
measurement on the system in state~$\state{S}$. On the basis of
the data obtained in~$n$ runs, one calculates a
posterior,~$p(\state{S})$, over state space, and then transforms
this posterior using the map~$\map{M}^*$, which is determined
by~$\map{M}$.

Although these two calculational routes cannot be expected to
agree for finite~$n$ owing to statistical fluctuations,
consistency requires that they agree~(so that the above diagram
commutes) in the limit as~$n \rightarrow \infty$.

\section{Deduction of the quantum formalism}
\label{sec:D}

In this section, we shall use the postulates described above to
derive the explicit form of the abstract quantum
model~$\model{q}(N)$, apart from the representation of temporal
evolution~(which is derived in Paper~II).  We shall also derive
the composite systems rule which allows the abstract quantum model
of a composite system to be related to the abstract quantum models
of its component systems.

The derivation will proceed as follows.
First, in Sec.~\ref{sec:D1}, we shall explore the consequences of
Postulate~2.3, the principle of information gain.  We shall find
that, if an information gain condition applies to a probabilistic
source with probability n-tuple~$\vect{P}= (P_1, P_2, \dots,
P_M)$~($M\geq 2$), then, if~$\vect{P}$ is represented as a unit
vector,~$\vect{Q}=(\sqrt{P}_1, \sqrt{P}_2, \dots, \sqrt{P}_M)$, in
a real `square-root of probability' space~(or $Q$-space), the
prior~$\Pr(\vect{Q}|\text{I})$ is uniform over the positive
orthant of the unit hypersphere in this space.

Second, following Postulates~1.1,~2.1, and~2.2, we shall represent
the state of a system,~$\state{S}(t)$, in a $2N$-dimensional
$Q$-space,~$Q^{2N}$.  We shall then use Postulate~2.4 to determine
the form of the function~$f$ that is introduced in the postulates.

Third, in Sec.~\ref{sec:D3}, we shall use Postulates~3, 3.1, 3.2,
3.3 and~4 in order to obtain a representation of physical
transformations of a system. We shall find that such
transformations can be represented by a subset of the orthogonal
transformations of the unit hypersphere in~$Q^{2N}$. We shall then
show that these transformations can, equivalently, be represented
by the set of unitary and antiunitary transformations of a
suitably-defined $N$-dimensional complex vector space.

Fourth, in Sec.~\ref{sec:D4}, we shall draw upon Postulate~1.2 in
order to obtain a representation of measurements on a system.

Fifth, in Sec.~\ref{sec:D5}, we shall use Postulate~5 to obtain a
rule, the \emph{composite system rule}, which determines the state
of a composite system in terms of the states of its sub-systems.

\subsection{Probabilistic Sources and Information Gain}
\label{sec:D1}

By postulates~1.1,~2.1 and~2.2, the measurement~$\mment{A}$ on the
system in state~$\state{S}(t)$ can, with respect to the outcomes
labeled~$i$ and~$a$ or~$b$, be modeled as the interrogation of a
$2N$-outcome probabilistic source with probability n-tuple
\begin{equation} \label{eqn:defn-of-P-lvect}
\lvect{P} = (P_1 P_{a|1}, P_1 P_{b|1}, \dots, P_N P_{a|N},
                                                    P_N P_{b|N}).
\end{equation}
From Postulate~2.2,~$f$ has range~$[-1, 1]$, so that all possible
values of~$\lvect{P}$ can be obtained by varying the state~$S(t)$.
 From Postulate~2.3, it therefore follows that, when this probabilistic source
 with any given~$\lvect{P}$ is
interrogated~$n$ times, the amount of Shannon-Jaynes information
obtained about~$\lvect{P}$ by an experimenter who does not know
the value of~$\lvect{P}$ is independent of~$\lvect{P}$ in the
limit as~$n \rightarrow \infty$. In order to implement this
condition, we shall begin by examining the process by which
information is gained about a probabilistic source.

\subsubsection{Information gain from a probabilistic source.}
\label{sec:D1-info-gain}

Consider an experiment in which an $M$-outcome probabilistic
source, with probability n-tuple~$\vect{P}=(P_1, P_2,\ldots,
P_M)$, is interrogated $n$~times, yielding the data string,~$D_n =
a_1a_2\ldots a_n$, of length~$n$, where~$a_r$ represents the value
of the~$r$th outcome~($r=1, \dots, n$).

Let us suppose that an experimenter knows that the data is
obtained from a probabilistic source, but does not the value
of~$\vect{P}$. Since the experimenter knows that the data is
generated by a probabilistic source, the order of the~$a_r$ is
irrelevant, the only relevant data being the number of
instances,~$m_i$ of each outcome,~$i$~($i=1, \dots, M$), which can
be encoded in the \emph{data n-tuple}~$\vect{m}=(m_1, m_2,\ldots,
m_M)$, or, equivalently, in the pair~$(\vect{f}, n)$,
where~$\vect{f} = \vect{m}/n$ is the frequency n-tuple.

The experimenter's knowledge about~$\vect{P}$ prior to the
experiment can be expressed as the prior probability density
function~$\Pr(\vect{P}|\text{I})$, where~I symbolizes the
knowledge that the experimenter possesses prior to performing the
interrogations.

After obtaining the data~$(\vect{f},n)$, the experimenter's state
of knowledge about~$\vect{P}$ is represented by the posterior
probability density function,~$\Pr(\vect{P}|\vect{f},n,\text{I})$.
The posterior can be related to the prior using Bayes' theorem,
\begin{equation} \label{eqn:posterior-probabilityP}
\Pr(\vect{P}| \vect{f}, n, \text{I}) = \frac{\Pr(\vect{f}|
\vect{P}, n, \text{I}) \Pr(\vect{P}| n, \text{I})}{ \Pr(\vect{f}|
n, \text{I}) },
\end{equation}
where the function~$\Pr(\vect{f}| \vect{P}, n, \text{I})$, known
as the likelihood, is given by
\begin{equation} \label{eqn:likelihood}
\Pr(\vect{f}| \vect{P}, n, \text{I}) = \frac{n!}{(nf_1)! \dots
(nf_M)!} P_1^{nf_1} \dots P_M^{nf_M}.
\end{equation}
The function~$\Pr(\vect{f}| n, \text{I})$ can be obtained from the
relation
\begin{equation} \label{eqn:normalisation}
\Pr(\vect{f}| n, \text{I}) = \idotsint_R
                        \Pr(\vect{f}| \vect{P}, n,
                        I) \Pr(\vect{P}|n, I) \,dP_1 \dots dP_N,
\end{equation}
where~$R$ is the set of~$\vect{P}$ satisfying the conditions~$0\le
P_i\le 1$~($i=1, \dots, N$) and~$\sum_i P_i = 1$.  In addition,
from Bayes' theorem,
\begin{equation}
\Pr(\vect{P}|n,\text{I}) \Pr(n|\text{I}) = \Pr(n|\vect{P},
\text{I}) \Pr(\vect{P}|\text{I}),
\end{equation}
and, using the fact that~$n$ is chosen freely by the experimenter
and therefore cannot depend upon~$\vect{P}$, which implies
that~$\Pr(n|\vect{P}, \text{I}) = \Pr(n|\text{I})$, it follows
that~$\Pr(\vect{P} |n, \text{I}) = \Pr(\vect{P} | \text{I})$.

In order to quantify the experimenter's change in knowledge
about~$\vect{P}$, we employ the Shannon-Jaynes information, which
is defined as follows.  First, the Shannon-Jaynes entropy
functional,
\begin{equation} \label{eqn:Shannon-Jaynes}
\sH \bigl[ F(\vect{P}) \bigr] = - \idotsint_R F(\vect{P})
                    \ln \frac{F(\vect{P})}{\Pr(\vect{P}|\text{I})} \, dP_1 \, dP_2 \dots
                    dP_N,
\end{equation}
is used to quantify the change in the experimenter's
uncertainty,~$\Delta H$, about~$\vect{P}$ as a result of obtaining
the data~$(\vect{f}, n)$.  The experimenter's gain of
Shannon-Jaynes information about~$\vect{P}$ is then defined
as~$\Delta K = -\Delta H$, which quantifies the decrease in the
experimenter's uncertainty~(equivalently, the increase in the
experimenter's knowledge) about~$\vect{P}$ as a result of
obtaining the data~$(\vect{f}, n)$.
The experimenter's gain of information about~$\vect{P}$ is
therefore given by
\begin{equation}        \label{eqn:Information-Gain}
\begin{split}
    \Delta K    &= \text{(Initial uncertainty about~\vect{P})}  \\
                   &\quad\quad\quad\quad\quad -\text{(Final uncertainty about~\vect{P})} \\
                &= \sH \bigl[ \Pr(\vect{P}  | \text{I}) \bigr]
                - \sH \bigl[ \Pr(\vect{P}  | \vect{f},n,\text{I}) \bigr]           \\
            &=  \idotsint_R \Pr(\vect{P}|  \vect{f},n,\text{I})
                    \ln \frac{\Pr(\vect{P}| \vect{f},n,\text{I})}{\Pr(\vect{P}|\text{I})} \, dP_1 \dots dP_N,  \\
\end{split}
\end{equation}
where we have used the fact that~$\sH\bigl[ \Pr(\vect{P}  |
\text{I}) \bigr] = 0$.

From this expression, one can see that, for given~$\Pr(\vect{P}  |
\vect{f},n,\text{I})$, the value of~$\Delta K$ depends upon the
prior probability,~$\Pr(\vect{P}| \text{I})$.  However, this prior
is left undetermined by the theory of probability. For
concreteness, consider the case where~$M=2$.  In that case,
the \emph{likelihood} is given by
\begin{equation}
\Pr(\vect{f}| \vect{P},n, \text{I}) =  \frac{n!}{m_1! (n-m_1)!}
P_1^{m_1} (1-P_1)^{n-m_1},
\end{equation}
which, in the limit of large~$n$, becomes very sharply peaked
around~$m_1=nP_1$ so that, in Eq.~\eqref{eqn:normalisation}, the
prior probability,~$\Pr(\vect{P}|\text{I})$, factors out of the
integrand, which, from Eq.~\eqref{eqn:posterior-probabilityP},
implies that the posterior~$\Pr(\vect{P}|\vect{f}, n,\text{I})$
can be approximated by
\begin{equation}
\Pr(\vect{P}| \vect{f}, n, \text{I}) = \frac{\Pr(\vect{f}|
\vect{P}, n, \text{I})}{\idotsint_R \Pr(\vect{f}| \vect{P}, n,
\text{I}) \,dP_1 \dots dP_N}.
\end{equation}
Consequently, the posterior~$\Pr(P_1| \vect{f}, n, \text{I})$ can
be approximated by a Gaussian function of variance~$\sigma^2 =
f_1(1-f_1)/n$.

For the purpose of illustration, suppose the prior
probability~$\Pr(\vect{P}|\text{I})$ is chosen to be uniform
on~$\sum_i P_i =1$, so that~$\Pr(P_1|\text{I}) = 1$. Then
Eq.~\eqref{eqn:Information-Gain} becomes
\begin{equation} \label{eqn:Information-Gain-Uniform-Prior}
\begin{split}
   \Delta K &=  \int \Pr(P_1|\vect{f}, n, \text{I})
                     \ln \frac{\Pr(P_1|\vect{f}, n,
                     I)}{\Pr(P_1|\text{I})} \, dP_1\\
            &=  \int \Pr(P_1|\vect{f}, n, \text{I})
                     \ln \Pr(P_1|\vect{f}, n,
                     I) \, dP_1 \\
            &\quad\quad\quad\quad\quad\quad - \int \Pr(P_1|\vect{f}, n, \text{I})
                     \ln \Pr(P_1|\text{I}) \, dP_1 \\
            &= - \ln  (\sigma \sqrt{2 \pi e})  \\
            & = \frac{1}{2} \ln \left(\frac{n}{2 \pi e}\right)
        - \frac{1}{2} \ln \left( f_1 \left(1-f_1 \right) \right),
\end{split}
\end{equation}
where we have made use of the standard result that, for a
Gaussian~$G_{\mu, \sigma}(x)$ over~$x$, with mean~$\mu$ and
standard deviation~$\sigma$, the integral
\begin{equation} \label{eqn:gaussian-entropy}
-\int_{-\infty}^{\infty} G_{\mu, \sigma}(x) \ln G_{\mu, \sigma}(x)
\,dx = \ln(\sigma \sqrt{2\pi e}).
\end{equation}
Equation~\eqref{eqn:Information-Gain-Uniform-Prior} clearly shows
that the value of~$\Delta K$ is dependent upon $f_1$. In the limit
of large~$n$,~$f_1$ tends to~$P_1$.  Thus, with the above choice
of the prior, the amount of information that the data provides
about~$\vect{P}$ depends upon the value of~$\vect{P}$. This
observation raises the possibility that one may be able to
choose~$\Pr(\vect{P}|\text{I})$ in such a way that~$\Delta K$ is
independent of~$P_1$ in the limit as~$n \rightarrow \infty$.

Let us then suppose that an $M$-outcome probabilistic source has a
prior~$Pr(\vect{P}|\text{I})$ such that the following condition
holds:

\medskip \noindent \emph{Information Gain Condition}. The amount of
Shannon-Jaynes information obtained about~$\vect{P}$ in~$n$
interrogations is independent of~$\vect{P}$ for all~$\vect{P}$.

\medskip\noindent In order to implement this condition, we can make use
of the fact the Shannon-Jaynes entropy is invariant under a change
of variables~\cite{Jaynes63}.  To illustrate the essential idea
underlying the implementation, we shall first give a simplified
argument for the case where~$M=2$; a more rigorous and general
argument is given in the appendix.

\medskip
\paragraph{Simplified argument for case~$M=2$.}

Suppose that~$\vect{P} = (P_1, P_2)$ is parameterized by the
parameter~$\lambda_1$, where the parametrization is bijective over
some interval,~$[\lambda_1^{(1)}, \lambda_1^{(2)}]$, of~$\lambda_1$,
and is differentiable.   Let us set~$\Pr(\lambda_1|\text{I})$ equal
to a constant~(fixed by normalization) over~$[\lambda_1^{(1)},
\lambda_1^{(2)}]$, and zero otherwise.

As stated above, in the limit of large~$n$, the
posterior~$\Pr(P_1|\text{I})$ takes the form of a Gaussian with
mean~$f_1$ and standard deviation~$\sigma$. Similarly, as we shall
later show explicitly, the
posterior~$\Pr(\lambda_1|\vect{f},n,\text{I})$ in this limit also
takes the form of a Gaussian distribution, with
mean~$\lambda_1^{(0)}$ defined through the relation~$f_1 =
P_1(\lambda_1^{(0)})$. To find the standard deviation,~$\sigma'$,
of the posterior over~$\lambda_1$, we use the relation~$P_1 =
P_1(\lambda_1)$,
    \begin{equation}
    \delta P_1 = \left(\frac{dP_1}{d\lambda_1}\right)\delta\lambda_1,
    \end{equation}
so that
    \begin{equation} \label{eqn:sigma-dash}
    \sigma' = \left|\frac{dP_1}{d\lambda_1}\right|^{-1} \sigma.
    \end{equation}
Using the expression for~$\sigma'$, the gain of information
about~$\lambda_1$~(and hence about~$\vect{P}$) is given by
\begin{equation}
\begin{split}
\Delta K &=  \int \Pr(\lambda_1|\vect{f}, n, \text{I})
                     \ln \frac{\Pr(\lambda_1|\vect{f}, n,
                     \text{I})}{\Pr(\lambda_1|\text{I})} \, d\lambda_1\\
            & = \int \Pr(\lambda_1|\vect{f}, n, \text{I})
                     \ln \Pr(\lambda_1|\vect{f}, n,
                     \text{I}) \, d\lambda_1 \\
            &\quad\quad\quad\quad\quad - \int \Pr(\lambda_1|\vect{f}, n, \text{I})
                     \ln \Pr(\lambda_1|\text{I}) \, d\lambda_1\\
            &= - \ln (\sigma' \sqrt{2\pi e}) - \ln\left(\Pr(\lambda_1|\text{I})\right) \\
            &= \ln\left[ \left|\frac{dP_1}{d\lambda_1}\right|
            \frac{1}{\sqrt{f_1(1-f_1)}} \right]  \\
                &\quad\quad\quad\quad\quad + \frac{1}{2}\ln\left(\frac{n}{2\pi e}\right) -
                \ln\left( \Pr(\lambda_1|\text{I}) \right).
\end{split}
\end{equation}
From this expression, one can see that the information gain will be
independent of~$\lambda_1$~(and therefore independent of~$P_1$) in
the limit as~$n \rightarrow \infty$ if and only if
\begin{equation} \label{eqn:condition-on-dP-dlambda}
\left|\frac{dP_1}{d\lambda_1}\right| \frac{1}{\sqrt{P_1(1-P_1)}} =
2a,
\end{equation}
where~$a$ is a real constant and is non-zero
since~$P_1(\lambda_1)$ is invertible, which implies that
\begin{equation} \label{eqn:P-1-cos-squared}
P_1 = \cos^2 \left(a\lambda_1 +b\right),
\end{equation}
where~$b$ is some real constant. Finally, from that fact
that~$\Pr(\lambda_1|\text{I})$ is a constant, using the relation
\begin{equation}
\Pr(P_1|\text{I}) |dP_1| = \Pr(\lambda_1|\text{I}) |d\lambda_1|,
\end{equation}
one finds that
\begin{equation} \label{eqn:prior-over-P-1}
\Pr(P_1|\text{I}) = \frac{1}{\pi} \frac{1}{\sqrt{P_1(1-P_1)}}.
\end{equation}

Hence, the above argument leads to the conclusion that the
information gain condition is satisfied for the case where~$M=2$
if and only if the prior~$\Pr(P_1|\text{I})$ takes the above form.
Furthermore, from Eqs.~\eqref{eqn:sigma-dash}
and~\eqref{eqn:condition-on-dP-dlambda}, it follows from the
expression for~$\sigma$ that
\begin{equation} \label{eqn:sigma-dash-final}
\sigma' = \frac{1}{2a\sqrt{n}}.
\end{equation}
Hence, that posterior over~$\lambda_1$ takes the form of a
Gaussian distribution whose standard deviation is independent
of~$\lambda_1^{(0)}$ and hence independent of~$\vect{P}$.

These results can be represented visually as follows. Define~$Q_i
= \sqrt{P_i}$~($0\leq Q_i \leq 1$, $i=1, 2$), and take~$\vect{Q} =
(Q_1, Q_2)$ to be a vector in a two-dimensional real Euclidean
space.
Then, from Eq.~\eqref{eqn:P-1-cos-squared}, it follows that
\begin{equation} \label{eqn:Q-1-cos}
Q_1 = \cos \left(a\lambda_1 +b\right).
\end{equation}
If we parameterize~$\vect{Q}$ as
\begin{equation} \label{eqn:Q-parameterisation}
\vect{Q} = (\cos\theta, \sin\theta),
\end{equation}
with~$\theta \in [0,\pi/2]$, we obtain that~$\theta = a\lambda_1
+b$. Since~$\Pr(\lambda_1|\text{I})$ is a constant, it follows
from the relation
\begin{equation} \label{eqn:lambda-theta-reln}
\Pr(\lambda_1|\text{I}) |d\lambda_1| = \Pr(\theta|\text{I})
|d\theta|
\end{equation}
that~$\Pr(\theta|\text{I})$ is also a constant.  Hence, the prior
over~$\theta$ is uniform over~$[0,\pi/2]$.  Conversely,
if~$\Pr(\theta|\text{I})$ is uniform, it follows from
Eq.~\eqref{eqn:Q-parameterisation} that the prior over~$P_1$ is
that given in Eq.~\eqref{eqn:prior-over-P-1}. Hence, the statement
that the prior over~$P_1$ is that given in
Eq.~\eqref{eqn:prior-over-P-1} is equivalent to the statement that
the prior is uniform over the positive quadrant of the unit circle
in~$Q^2$.

We note also that, from Eq.~\eqref{eqn:sigma-dash-final}, using
the relation~$\theta = a\lambda_1 +b$ and
Eq.~\eqref{eqn:lambda-theta-reln}, it follows that the
posterior,~$\Pr(\theta|\vect{f}, n, \text{I})$, over~$\theta$
takes the form of a Gaussian with standard
deviation~$\sigma_{\theta} = 1/2\sqrt{n}$.

\medskip
\paragraph{Statement of the general result.}

As shown in the appendix, the above results for~$M=2$ generalize
as follows.   For an $M$-outcome probabilistic source, the
information gain condition is satisfied if and only if
\begin{equation} \label{eqn:general-prior-over-P}
\Pr(\vect{P}|\text{I}) = \frac{2}{A_{M-1}}
\frac{1}{\sqrt{P_1\dots, P_M}} \,\,\delta\left(1 - \sum_i
P_i\right),
\end{equation}
where~$A_{M-1}$ is the surface area of a unit~$M$-ball.

Consider an $M$-dimensional real Euclidean space,~$Q^M$, with
axes~$Q_1, Q_2, \dots, Q_M$.  If we define the vector~$\vect{Q} =
(Q_1, Q_2, \dots, Q_M)$ such that~$Q_i = \sqrt{P_i}$, where~$0\leq
Q_i \leq 1$, then every~$\vect{Q}$ that represents a probability
n-tuple lies on the positive orthant,~$\orthant$, of the unit
hypersphere,~$\hypersphere$. Then, using the relation
\begin{equation} \label{eqn:Q-P-transformation}
\Pr(\vect{Q}|\text{I}) = \left| \frac{\partial(P_1, \dots,
P_M)}{\partial(Q_1, \dots, Q_M)}\right| \Pr(\vect{P}|\text{I}),
\end{equation}
it follows that the prior over~$\vect{Q}$ is given by
\begin{equation} \label{eqn:general-prior-over-Q}
\Pr(\vect{Q}|\text{I}) = \frac{2^{M+1}}{A_{M-1}} \,\,\delta\left(1
- |\vect{Q}|^2\right),
\end{equation}
which implies that the prior is uniform over~$\orthant$.
Conversely, if the prior is uniform over~$\orthant$, it follows
that the prior over~$\vect{P}$ is that given in
Eq.~\eqref{eqn:general-prior-over-P}.  Finally,  in the limit
as~$n \rightarrow \infty$, the posterior over~$\orthant$ is a
symmetric Gaussian with standard deviation~$1/2\sqrt{n}$.

\subsubsection{Prior Probabilities over~$\lvect{P}$}
\label{sec:D1-prior-probabilities-in-state-space}

From the above discussion, it follows that Postulate~2.3 imposes a
particular prior over~$\lvect{P}$~(see
Eq.~\eqref{eqn:defn-of-P-lvect}), namely
\begin{equation} \label{eqn:prior-over-lvectP}
\Pr(\lvect{P}|\text{I}) = \frac{2}{A_{2N-1}}
                \frac{1}{\sqrt{\tilde{P}_1\dots \tilde{P}_{2N}}}
                \,\,\delta \left(1- \sum_{q=1}^{2N}
                \tilde{P}_q\right),
\end{equation}
where~$\tilde{P}_q$ denotes the~$q$th component of~$\lvect{P}$.
As in the previous section, we shall describe~$\lvect{P}$ as a
unit vector,
\begin{equation} \label{eqn:defn-of-lvect-Q}
\lvect{Q} = (Q_1, Q_2, \dots, Q_{2N})
\end{equation}
in~$Q^{2N}$, where~$Q_q = \sqrt{\tilde{P}_q}$ and~$0\le Q_q \le 1$.



From the results of the previous section, the prior over the
positive orthant of the unit hypersphere is uniform and, after
obtaining the data from~$n$ runs of the experiment, in the limit
as~$n \rightarrow \infty$, the posterior can be represented by a
symmetric Gaussian distribution over the positive orthant, with
standard deviation~$1/2\sqrt{n}$.

\subsubsection{Determination of function~$f$}
\label{sec:D1-determining-f}

In order to determine the unknown function~$f$ which is introduced
in Postulate~2.2, we shall first use the prior over~$\lvect{P}$ to
determine the priors~$\Pr(P_{a|i}|\text{I})$~($i=1,\dots, N)$, and
then use the relationship~$P_{a|i} = F(\var_i)$, where~$F(\var_i)
= f^2(\var_i)$~(Postulate~2.2) and the uniformity of the
prior~$\Pr(\var_i|\text{I})$~(Postulate~2.4) to determine~$f$.

To determine the prior~$\Pr(P_{a|i}|\text{I})$, the first step is
to find the prior~$\Pr(P_1, P_{a|1}, \dots, P_N, P_{a|N})$ using
the prior in Eq.~\eqref{eqn:prior-over-lvectP}, where, from
Eq.~\eqref{eqn:defn-of-P-lvect}, and using the fact that~$P_{a|i}
+ P_{b|i} =1$,
\begin{gather}
\tilde{P}_{2i-1} = P_i P_{a|i} \\
\tilde{P}_{2i} =  P_i (1-P_{a|i}),
\end{gather}
for~$i=1, \dots, N$. Using the relation
\begin{multline} \label{eqn:1-2-parameterisation}
\Pr(P_1, P_{a|1}, \dots, P_N, P_{a|N}| \text{I}) = \\
\left|\frac{\partial(\tilde{P}_1,\tilde{P}_2,\dots,
                       \tilde{P}_{2N-1}, \tilde{P}_{2N})}{\partial(P_1, P_{a|1}, \dots,P_N, P_{a|N})}\right|
                            \Pr(\lvect{P}| \text{I}),
\end{multline}
in which the modulus of Jacobian evaluates to~$\prod_i P_i$, we
find
\begin{multline} \label{eqn:prior-over-1-I}
\Pr(P_1, P_{a|1}, \dots, P_N, P_{a|N}| \text{I}) =
                \frac{2}{A_{N-1}} \\
                \times \prod_{i=1}^{N}
                \frac{1}{\sqrt{P_{a|i}(1-P_{a|i})}}
                \,\,\delta\left(1 -  \sum_{i=1}^{N} P_i \right)
\end{multline}

Next, to find the marginal probability over~$P_{a|i}$, we first
marginalize over~$P_1, \dots, P_N$, to obtain
\begin{equation}
\Pr(P_{a|1}, \dots, P_{a|N}| \text{I})
                =
                \prod_{i=1}^{N}
                \frac{1}{\pi}
                \frac{1}{\sqrt{P_{a|i}(1-P_{a|i})}},
\end{equation}
and then marginalize over~$P_{a|1}, \dots, P_{a|i-1}, P_{a|i+1},
\dots, P_{a|N}$, to obtain
\begin{equation} \label{eqn:prior-over-P-a-sub-i}
\Pr(P_{a|i}| \text{I}) = \frac{1}{\pi}
                    \frac{1}{\sqrt{P_{a|i}(1-P_{a|i})}}.
\end{equation}

From Postulate~2.2, the probability~$P_{a|i} = F(\var_i)$, and,
from Postulate~2.4, the prior~$\Pr(\var_i|\text{I})$ is uniform.
Using Eq.~\eqref{eqn:prior-over-P-a-sub-i} and the relation
\begin{equation}
\Pr(P_{a|i}|\text{I})|dP_{a|i}| \propto
\Pr(\var_i|\text{I})|d\var_i|,
\end{equation}
where the proportionality is due to the fact that the
prior~$\Pr(\var_i|\text{I})$ is non-normalizable, it follows that
\begin{equation}
\frac{dF(\var_i)}{d\var_i} \propto
\sqrt{F(\var_i)\left(1-F(\var_i)\right)},
\end{equation}
which has the general solution
\begin{equation} \label{eqn:F-function}
F(\var_i) = \cos^2(a\var_i + b),
\end{equation}
where~$a$ and~$b$ are real constants, and where~$a\neq 0$ since,
by Postulate~2.2, the function~$f(\var_i)$ is not a constant
function.  Hence, the functions~$f$ and~$\tilde{f}$~(see
Postulate~2.2) have the form
\begin{equation} \label{eqn:f-function}
\begin{aligned}
 f(\var_i) &= \pm \cos(a\var_i +b)  \\
 \tilde{f}(\var_i) &= \pm \sin(a\var_i +b), \\
\end{aligned}
\end{equation}
where the signs of~$f$ and~$\tilde{f}$ are undetermined.

\subsubsection{Representation of state space.}
\label{sec:state-space}

Above, we have represented~$\lvect{P}$ as a unit
vector,~$\lvect{Q}$, on the positive orthant of the unit hypersphere
in~$Q^{2N}$.  Now, the binary-valued degrees of freedom in~$S(t)$
described in Postulate~2.2 are encoded into the signs of
the~$Q_{a|i}$ and~$Q_{b|i}$.  Therefore, if we remove the condition
of positivity imposed on the~$Q_q$, then, given~$\lvect{Q}$ on the
unit hypersphere,~$S^{2N-1}$, the probabilities~$\tilde{P}_q$ can be
read out using the relation~$\tilde{P}_q = Q_q^2$, and the values of
the $2N$~binary degrees of freedom are read out from the
$2N$~signs~(either~$+$ or~$-$) of the~$Q_q$. Graphically, the
orthant containing~$\lvect{Q}$ encodes the values of the binary
degrees of freedom, while the location of~$\lvect{Q}$ within a given
orthant encodes the values of the~$\tilde{P}_q$.

According to Postulate~2.2,~$\lvect{P}$ and the values of the
$2N$~binary degrees of freedom constitute all of the information
that the quantum state,~$S(t)$, of the system provides about
objectively realized physical events when measurement~$\mment{A}$
is performed on the system.  Therefore, the value of~$\lvect{P}$
and the values of the binary degrees of freedom can be taken to
completely represent~$S(t)$ with respect to
measurement~$\mment{A}$.

In particular,~$\lvect{Q}$ in~$S^{2N-1}$ represents the
state~$S(t)$, where now the only condition imposed on the~$Q_q$ is
that~$\tilde{P}_q = Q_q^2$ for~$q=1,\dots, 2N$. Hence, the
set,~$S^{2N-1}$, of unit vectors in~$Q^{2N}$ represents the state
space of the system.

Using the functions~$f$ and~$\tilde{f}$ from
Eqs.~\eqref{eqn:f-function}, taking~$a=1$ and~$b=0$ and choosing
the positive signs, we can write~$Q_{a|i} = \cos\var_i$
and~$Q_{b|i} = \sin\var_i$, and therefore can write the state of a
system with respect to measurement~$\mment{A}$ as
\begin{equation} \label{eqn:defn-of-lvect-Q-2}
\begin{split}
\lvect{Q}   &= (\sqrt{P_1} Q_{a|1}, \sqrt{P_1} Q_{b|1}, , \dots,
                            \sqrt{P_N} Q_{b|N}) \\
            &= (\sqrt{P_1} \cos\var_1, \sqrt{P_1} \sin\var_1, \ldots,
            \sqrt{P_N} \sin\var_N).
\end{split}
\end{equation}
In Paper~II, we shall show that the above choice of the positive
signs for the functions~$f$ and~$\tilde{f}$ and choice of the
constants~$a,b$ involves no loss of generality.

The prior over~$S^{2N-1}$ is the product of the priors due to the
binary degrees of freedom and due to~$\lvect{P}$. Since~$Q_1 =
\sqrt{P}_1 \cos\var_1$ and~$\Pr(\var_1|\text{I})$ is uniform, it
follows that the sign of~$Q_1$ is \textit{a priori} equally likely
to be positive or negative, and similarly for~$Q_2, \dots, Q_{2N}$.
Therefore, each orthant is, \textit{a priori}, equally likely to
contain~$\lvect{Q}$. Since the prior due to~$\lvect{P}$ is expressed
by a uniform prior over the positive orthant, the resultant prior
over~$S^{2N-1}$ is uniform.

In the case of the posterior over~$S^{2N-1}$, the orthant
containing~$\lvect{Q}$ is known with a probability very close to
unity in the limit of large~$n$.  Therefore, the posterior
over~$S^{2N-1}$ in the limit as~$n\rightarrow \infty$ is
arbitrarily well approximated by a probability density function
that consists of a symmetric Gaussian in the orthant
containing~$\lvect{Q}$, and is zero in all other orthants.

\subsection{Mappings}
\label{sec:D3}

According to Postulate~3, a physical transformation of a physical
system is represented by a map,~$\map{M}$, from state space to
itself. In this section, the general form of mappings that are
consistent with the postulates will be determined.

The derivation will be based upon Postulates~3.1--3.3 and
Postulate~4, and will proceed in four steps:
\begin{enumerate}
    \item[(1)] Show that Postulates~3.1 and~4 imply that~$\map{M}$ is
    an orthogonal transformation
    of the unit hypersphere in~$Q^{2N}$.

    \item[(2)] Show that the imposition of Postulate~3.2 restricts~$\map{M}$
    to a subset of the set of orthogonal transformations, and that these
    transformations can be recast as unitary or antiunitary transformations
    acting on a suitably-defined complex vector space.

    \item[(3)] Show that any unitary or antiunitary transformation represents
    an orthogonal transformation satisfying
    Postulates~3.1,~3.2, and~4.

    \item[(4)] Show that a physical transformation which depends continuously
    upon a real-valued parameter n-tuple can be represented by either
    unitary or antiunitary transformations, that a
    continuous physical transformation can only be represented by
    unitary transformations, and that a discrete transformation
    can be represented by either a unitary or an antiunitary
    transformation.
\end{enumerate}

\subsubsection{Step 1: Orthogonal Transformations}

As discussed in Sec.~\ref{sec:state-space}, the state space of a
system can be represented by the set of unit vectors,~$S^{2N-1}$,
in the~$2N$--dimensional space~$Q^{2N}$.  According to
Postulate~3.1, the map~$\map{M}$ over state space is one-to-one.
Hence, the map over~$S^{2N-1}$, which we shall denote
by~$\map{T}$, is one-to-one.

We can now impose two further constraints on~$\map{T}$. First, we
have found that the prior,~$\Pr(\lvect{Q}|\text{I})$, is uniform
over the unit hypersphere. Under map~$\map{T}$, the prior
transforms into the probability density
function,~$\tilde{p}(\lvect{Q}')$, given by
\begin{equation} \label{eqn:general-T-action}
\tilde{p}(\lvect{Q}') = \Pr(\lvect{Q}|\text{I})
                \left|\frac{\partial(Q_1', \dots,
                Q_{2N}')}{\partial(Q_1, \dots,
                Q_{2N})}
                \right|^{-1},
\end{equation}
where~$\lvect{Q}' = \map{T}(\lvect{Q})$, with~$\lvect{Q}= (Q_1,
\dots, Q_{2N})$ and~$\lvect{Q}'= (Q_1', \dots, Q_{2N}')$. However,
under the physical transformation represented by~$\map{T}$, no
measurement has been performed by the experimenter and therefore the
prior assigned by the experimenter over the unit hypersphere must
remain unchanged. That is, the map,~$\map{T}$ must be such
that~$\tilde{p}(\lvect{Q}')$ is also uniform over the unit
hypersphere, which implies that
\begin{equation} \label{eqn:Jacobian-constancy}
                \left|\frac{\partial(Q_1', \dots,
                Q_{2N}')}{\partial(Q_1, \dots,
                Q_{2N})}
                \right|
                = 1.
\end{equation}
Hence, in general, under~$\map{T}$, the probability density
function~$p(\lvect{Q})$  transforms to the probability density
function
\begin{equation} \label{eqn:final-T-action}
\tilde{p}(\lvect{Q}') = p(\lvect{Q}).
\end{equation}

Second, from Postulate~4, we can, in the limit as~$n \rightarrow
\infty$, obtain a posterior over~$Q^{2N}$ of a system in
state~$\lvect{Q}' = \map{T}(\lvect{Q})$ in one of two equivalent
ways:
\begin{itemize}
    \item[(i)] perform measurement~$\mment{A} \in \set{A}$ upon~$n$
    copies of a system in state~$\lvect{Q}$, and then use~$\map{T}$ to
    transform the posterior~$\Pr(\lvect{Q}|D_n,\text{I})$ based on the
    data,~$D_n$, consisting of the realized outcomes, or
    \item[(ii)] perform measurement~$\mment{A} \in \set{A}$
    upon~$n$ copies of a
    system in state~$\lvect{Q}'$, and write down the
    posterior~$\Pr(\lvect{Q}|D_n',\text{I})$ based on the
    data,~$D_n'$, consisting of the realized outcomes.
\end{itemize}

Now, from the discussion of Sec.~\ref{sec:D1-info-gain}, in the
limit as~$n\rightarrow \infty$, the posterior, which we shall denote
by~$h$, over the unit hypersphere in~$Q^{2N}$, is zero apart from in
one orthant, where it takes the form of a symmetric Gaussian
function whose standard deviation is a function of~$n$ only.
Therefore, the posteriors~$\Pr(\lvect{Q}|D_n,\text{I})$
and~$\Pr(\lvect{Q}|D_n',\text{I})$ are both of this form, with the
symmetric Gaussian functions having the same standard deviation. In
order that Postulate~4 holds for any measurement~$\mment{A} \in
\set{A}$ and for any possible interaction in~$\set{I}$, it therefore
follows that, in addition to satisfying
Eq.~\eqref{eqn:final-T-action}, the map~$\map{T}$ must satisfy the
condition that any probability density function of the form~$h$,
containing a symmetric Gaussian with given standard deviation, is
mapped to a probability density function which is asymptotically
equal to a probability density function of the form~$h$ that
contains a symmetric Gaussian with the same standard deviation.

One can readily see that any orthogonal transformation of the unit
hypersphere will satisfy this condition since such a
transformation will take a symmetric Gaussian with given standard
derivation to another symmetric Gaussian with the same standard
derivation.  We shall now show that, in fact, the set of
all~$\map{T}$ is precisely equal to the set of orthogonal
transformations over~$S^{2N-1}$

First, we shall show that, in order to satisfy the above condition,
the map~$\map{T}$ must preserve the distance between any two points
that lie in the same orthant on the unit hypersphere. To see this,
consider the converse. Suppose, then, that there exist two
points,~$\lvect{Q}_1, \lvect{Q}_2$ on the same orthant of the
hypersphere such that~$\metricd(\lvect{Q}_1, \lvect{Q}_2) \neq
\metricd(\lvect{Q}_1', \lvect{Q}_2')$ where primes indicate vectors
transformed by~$\map{T}$, and where~$\metricd(\lvect{Q}_1,
\lvect{Q}_2)$ denotes the distance between~$\lvect{Q}_1$
and~$\lvect{Q}_2$ according to some given distance
function,~$\metricd$. Choose a function~$h$ containing a symmetric
Gaussian function which peaks at~$\lvect{Q}_1$, and define the
set~$\set{Q}^{(r)}$ as the set of all points in the orthant at a
distance~$r = \metricd(\lvect{Q}_1, \lvect{Q}_2)$
from~$\lvect{Q}_1$.

Since the Gaussian is symmetric about~$\lvect{Q}_1$,~$h(\lvect{Q}_a)
= h(\lvect{Q}_b)$ for all~$\lvect{Q}_a, \lvect{Q}_b \in
\set{Q}^{(r)}$.  Therefore,~$\set{Q}^{(r)}$ is a subset of a
$2(N-1)$--spherical equiprobability contour centered
around~$\lvect{Q}_1$ of radius~$r$.  Since~$h(\lvect{Q}_2) -
h(\lvect{Q}_1)$ decreases monotonically with~$\metricd(\lvect{Q}_1,
\lvect{Q}_2)$,~$\set{Q}^{(r)}$ contains \emph{all} the points in the
orthant with the value~$g(\lvect{Q}_2)$.

Under the mapping~$\map{T}$, the points~$\lvect{Q}_1', \lvect{Q}_2'$
are such that~$\tilde{h}(\lvect{Q}_1') = h(\lvect{Q}_1)$
and~$\tilde{h}(\lvect{Q}_2') = h(\lvect{Q}_2)$, where~$\tilde{h}$ is
the transformed posterior, so that~$\set{Q}^{(r)}$ maps to the
equiprobability contour~$\set{Q}'^{(r)}$.  Now, by
assumption,~$\map{T}$ maps~$h$ onto a function,~$\tilde{h}$, that
asymptotically approaches a probability density function of the same
form as~$h$. Therefore, in particular,~$\map{T}$ must preserve the
shape of the Gaussian function and its equiprobability contours.
However, it was supposed that~$\metricd(\lvect{Q}_1', \lvect{Q}_2')
\neq r$. Therefore,~$\set{Q}'^{(k)}$ contains a
point,~$\lvect{Q}'_2$, that is not a distance~$r$
from~$\lvect{Q}_1$.  Therefore, unlike~$\set{Q}^{(r)}$, the
set~$\set{Q}'^{(r)}$ is not a subset of a~$2(N-1)$--spherical
equiprobability contour of radius~$r$, which leads to a
contradiction.  Therefore, the original supposition must be false,
which implies that~$\map{T}$ preserves the distance between any two
points~$\lvect{Q}_1, \lvect{Q}_2$ that lie in the same orthant of
the hypersphere.

In the case of two points that lie in different orthants, we argue
as follows.  Consider first the simplest case where two
points,~$\lvect{Q}_1, \lvect{Q}_2$, lie in adjacent orthants
and~$N=2$.  Now, choose two points~$\lvect{Q}_1', \lvect{Q}_2'$,
that lie in the first and second orthants, respectively.  From the
above result, the distances~$\metricd(\lvect{Q}_1, \lvect{Q}_1')$
and~$\metricd(\lvect{Q}_2, \lvect{Q}_2')$ are preserved
under~$\map{T}$. Suppose now that the points~$\lvect{Q}_1',
\lvect{Q}_2'$ are brought closer together, whilst still remaining in
their respective orthants.  In the limit as~$\metricd(\lvect{Q}_1',
\lvect{Q}_2') \rightarrow 0$ such that~$\lvect{Q}_1', \lvect{Q}_2'$
tend to the point~$\lvect{Q}'$ that lies on the boundary between the
two orthants, it follows that the distances~$\metricd(\lvect{Q}_1,
\lvect{Q}')$ and~$\metricd(\lvect{Q}_2, \lvect{Q}')$ are preserved
under~$\map{T}$.

Similarly, one can choose two further pairs of
points,~$\lvect{Q}_1'', \lvect{Q}_2''$ and~$\lvect{Q}_1''',
\lvect{Q}_2'''$, that lie in the first and second octants
respectively, and conclude that, if they tend to the
points~$\lvect{Q}'', \lvect{Q}'''$, respectively, which both lie on
the boundary between the two orthants, the
distances~$\metricd(\lvect{Q}_i, \lvect{Q}'')$
and~$\metricd(\lvect{Q}_i, \lvect{Q}''')$,for~$i=1, 2$, are also
preserved under~$\map{T}$. Let us now choose~$\lvect{Q}',
\lvect{Q}'', \lvect{Q}'''$ to be distinct points. Since the
distances of~$\lvect{Q}_1$ and~$\lvect{Q}_2$ from~$\lvect{Q}',
\lvect{Q}'', \lvect{Q}'''$ are all invariant under~$\map{T}$, it
follows that the distance~$\metricd(\lvect{Q}_1, \lvect{Q}_2)$ is
invariant.

The above argument can be readily generalized to the case of two
points in adjacent orthants for general~$N$, and, further, to the
case where two points are in non-adjacent orthants.

Second, since~$\map{T}$ preserves the distance between any two
points on the hypersphere, it is an orthogonal transformation
of~$S^{2N-1}$.  But we have already noted that any orthogonal
transformation of~$S^{2N-1}$ is an acceptable map~$\map{T}$.
Hence, the set of all~$\map{T}$ is equal to the set of orthogonal
transformations of~$S^{2N-1}$.

\subsubsection{Step 2: Imposition of Postulate~3.2}
\label{sec:M-is-equivalent-to-unitary}

Postulate~3.2 requires that the outcome probabilities~$P_1', P_2',
\dots, P_N'$ of measurement~$\mment{A}$ performed on a system in
state~$\lvect{Q}' = \map{T}(\lvect{Q})$ are unaffected if, in the
state~$\lvect{Q}$ written down with respect to
measurement~$\mment{A}$, an arbitrary real constant,~$\var_0$, is
added to each of the~$\var_i$.

Since~$\map{T}$ is an orthogonal transformation, it can be
represented by the~$2N$--dimensional orthogonal matrix,~$M$. Under
its action, the vector~$\lvect{Q}$ transforms as
\begin{equation} \label{eqn:Q-dash-is-Q-rotated}
    \lvect{Q}' = \rmatrix{M} \lvect{Q}.
\end{equation}
Multiplying this out, the form of~$P_k'$ in terms of the~$P_i$
and~$\var_i$ is
%
\begin{equation} \label{eqn:basic-P-k-dash}
\begin{split}
 P_k'   &= \sum_i   P_i  \big[
                        (M_{2k-1, 2i-1}\cos\var_i + M_{2k-1,
                        2i}\sin\var_i)^2 \\
        &\quad\quad\quad\quad\quad+
                        (M_{2k, 2i-1}\cos\var_i + M_{2k, 2i}\sin\var_i)^2
                        \big] \\
        &\quad+
        2 \sum_{\substack{i,j \\ i<j}} \sqrt{P_i P_j}
            \begin{aligned}[t]
                &\big[A_{kij}\cos\var_i\cos\var_j
                                +B_{kij}\cos\var_i \sin\var_j \\
                &+C_{kij}\sin\var_i\cos\var_j
                                +D_{kij}\sin\var_i\sin\var_j\big],
            \end{aligned}
 \end{split}
 \end{equation}
where
  \begin{equation} \label{eqn:matrix-definitionsB}
    \begin{aligned}
     A_{kij} &= M_{2k-1, 2i-1} M_{2k-1, 2j-1} +  M_{2k, 2i-1} M_{2k, 2j-1}         \\
     B_{kij} &=  M_{2k-1, 2i-1}M_{2k-1, 2j} +  M_{2k, 2i-1} M_{2k, 2j}              \\
     C_{kij} &=M_{2k-1,2i}M_{2k-1, 2j-1} +  M_{2k, 2i} M_{2k, 2j-1}                 \\
     D_{kij} &=  M_{2k-1, 2i}M_{2k-1, 2j} + M_{2k, 2i} M_{2k, 2j}.
    \end{aligned}
    \end{equation}

In order to implement Postulate~3.2, it is helpful to rewrite the
above expression for~$P_k'$ so that the~$\var_i$ appear in the
form~$(\var_i \pm \var_j)$ since the value of terms of the
form~$(\var_i - \var_j)$ remains unchanged under the addition
of~$\var_0$ to each of the~$\var_i$.  One finds that
\begin{equation} \label{eqn:basic-P-k-dash-extra}
\begin{split}
 P_k'
    &= \frac{1}{2}
        \sum_i (\alpha_{ki} + \beta_{ki})P_i \\
        &\quad+     \sum_{\substack{i,j \\ i<j}} \sqrt{P_i P_j}
                    \big[
                    (A_{kij} + D_{kij})\cos(\var_i-\var_j) \\
                    &\quad\quad\quad\quad\quad\quad\quad
                    - (B_{kij} - C_{kij})\sin(\var_i-\var_j)
                    \big]                                                        \\
        &\quad+\sum_i
                   \cos(\var_i+\var_{i\oplus 1})
                    \bigg[
                        \frac{1}{2} (\alpha_{ki}-\beta_{ki})P_i
                            \cos(\var_i - \var_{i \oplus 1})\\
                        &\quad\quad\quad\quad\quad\quad\quad\quad\quad\quad\quad+
                         \gamma_{ki}P_i \sin(\var_i - \var_{i\oplus 1})
                    \bigg]   \\                                              \\
        &\quad+\sum_i \sin(\var_i+\var_{i\oplus 1})
                    \bigg[
                        -\frac{1}{2} (\alpha_{ki}-\beta_{ki})P_i
                            \sin(\var_i - \var_{i\oplus 1}) \\
                        &\quad\quad\quad\quad\quad\quad\quad\quad\quad\quad\quad+
                             \gamma_{ki}P_i \cos(\var_i - \var_{i\oplus 1})
                    \bigg] \\
        &\quad+\sum_{\substack{i,j \\ i<j}} \sqrt{P_i P_j}
           \big[
                (A_{kij} - D_{kij}) \cos(\var_i+\var_j) \\
                 &\quad\quad\quad\quad\quad\quad\quad\quad\quad+
                (B_{kij} + C_{kij}) \sin(\var_i+\var_j)
            \big]
    \end{split}
 \end{equation}
where
    \begin{equation}
    \begin{aligned} \label{eqn:matrix-definitionsA}
        \alpha_{ki} &= M_{2k-1, 2i-1}^2 + M_{2k, 2i-1}^2                           \\
        \beta_{ki} &= M_{2k-1, 2i}^2 + M_{2k,2i}^2                                 \\
        \gamma_{ki} &= M_{2k-1, 2i-1}M_{2k-1,2i} + M_{2k, 2i-1}M_{2k, 2i}
    \end{aligned}
    \end{equation}
and~$\oplus$ denotes addition modulo~$N$.

Postulate~3.2 must hold for any~$P_i$ and~$\var_i$.   Therefore,
in particular, it must be true for the special case where all but
one, say~$P_i$, of the~$P_i$ are zero and all of the~$\var_i$ have
the same value. In this case, Eq.~\eqref{eqn:basic-P-k-dash-extra}
simplifies to
\begin{equation} \label{eqn:basic-P-k-dash1}
    \begin{split}
 P_k'   &= \frac{1}{2} \left(\alpha_{ki} + \beta_{ki} \right) \\
        &\quad+
            \frac{1}{2} \left(\alpha_{ki}-\beta_{ki} \right)
             \cos(\var_i+\var_{i\oplus 1})
            +
            \gamma_{ki} \sin(\var_i+\var_{i\oplus 1}).
    \end{split}
 \end{equation}
We require that~$P_k'$ remains unchanged as a result of the
addition of any constant~$\var_0 \in \numberfield{R}$ to
the~$\var_i$. However, a linear combination of the
functions~$\cos(\var_i+\var_{i\oplus 1})$
and~$\sin(\var_i+\var_{i\oplus 1})$ in which at least one of the
coefficients is non-zero is zero only on a discrete set of points.
Therefore, the coefficients of the
functions~$\cos(\var_i+\var_{i\oplus 1})$
and~$\sin(\var_i+\var_{i\oplus 1})$ must vanish, so that the
conditions
\begin{equation}\label{eqn:rotation-matrix-constraintsA}
\alpha_{ki} = \beta_{ki} \quad \text{and} \quad \gamma_{ki} = 0
\quad \text{for all~$i,k$}
\end{equation}
must hold.

Consider now a second special case where two of the~$P_i$,
say~$P_i$ and~$P_j$~$(i\neq j)$ are set equal to~$1/2$, and the
remainder are set to zero.  Then, taking into account the above
conditions, Eq.~\eqref{eqn:basic-P-k-dash-extra} reduces to
\begin{equation} \label{eqn:basic-P-k-dash2}
    \begin{split}
 P_k'   &= \frac{1}{2} \left[
        \frac{1}{2}(\alpha_{ki} + \beta_{ki}) +  \frac{1}{2}(\alpha_{kj} + \beta_{kj})
                        \right] \\
            &\quad+ \frac{1}{2}
            \bigg[
                (A_{kij} + D_{kij})\cos(\var_i-\var_j) \\
                &\quad\quad\quad\quad\quad\quad-
                    (B_{kij} - C_{kij})\sin(\var_i-\var_j)
            \bigg]  \\
        &\quad+\frac{1}{2}
            \bigg[
                (A_{kij} - D_{kij}) \cos(\var_i+\var_j) \\
                &\quad\quad\quad\quad\quad\quad+
                (B_{kij} + C_{kij}) \sin(\var_i+\var_j)
            \bigg].
    \end{split}
 \end{equation}
Once again, in order that~$P_k'$ remains unchanged as a result of
the addition of~$\var_0 \in \numberfield{R}$ to the~$\var_i$, the
coefficients of the
functions~$\cos(\var_i+\var_j)$and~$\sin(\var_i+\var_j)$ must
vanish, so that a second set of conditions,
\begin{multline}
A_{kij} = D_{kij} \quad \text{and} \quad B_{kij} = - C_{kij} \\
\qquad \text{for all~$i, j$ and~$k$, with~$i \neq j$},
\label{eqn:rotation-matrix-constraintsB}
\end{multline}
must hold.

The most general matrix, $M$, which satisfies the first set of
conditions, expressed in
Eqs.~\eqref{eqn:rotation-matrix-constraintsA}, can be written in
the form of a~$N$-by-$N$ array of two-by-two sub-matrices,
\begin{equation} \rmatrix{M} =
\begin{pmatrix} \label{eqn:restricted-rotation}
     T^{(11)} &  T^{(12)} & \dots &
        T^{(1N)} \\
     T^{(21)} &  T^{(22)} & \dots &
         T^{(2N)} \\
    \hdotsfor[2]{4}                 \\
     T^{(N1)} &  T^{(N2)} & \dots &
         T^{(NN)}
\end{pmatrix},
\end{equation}
where
\[ T^{(ij)} = \sqrt{\alpha_{ij}}
        \begin{pmatrix}
          \cos \varphi_{ij}  & -\sigma_{ij}\sin\varphi_{ij} \\
          \sin \varphi_{ij}  &  \sigma_{ij}\cos\varphi_{ij}
        \end{pmatrix}
\]
is a two-by-two matrix composed of a enlargement matrix~(scale
factor~$\sqrt{\alpha_{ij}}$) and a rotation matrix if~$\sigma_{ij}
=1$ or a reflection-rotation matrix~(that is, a matrix
representing a reflection followed by rotation) if~$\sigma_{ij}
=-1$, with rotation angle~$\varphi_{ij}$ in either case.

In terms of the~$\sigma_{ij}$ and the~$\alpha_{ij}$,
Eqs.~\eqref{eqn:matrix-definitionsB} then becomes
\begin{equation}
\begin{aligned}
 A_{kij} &= \sqrt{\alpha_{ki}\alpha_{kj}} \left( \cos\varphi_{ki} \cos\varphi_{kj}
                                    + \sin\varphi_{ki} \sin\varphi_{kj} \right) \\
B_{kij} &= \sigma_{kj}\sqrt{\alpha_{ki}\alpha_{kj}} \left(
-\cos\varphi_{ki}\sin\varphi_{kj}
                                    + \sin\varphi_{ki} \cos\varphi_{kj} \right) \\
C_{kij} &= \sigma_{ki}\sqrt{\alpha_{ki}\alpha_{kj}} \left(
-\sin\varphi_{ki} \cos\varphi_{kj}
                                    + \cos\varphi_{ki} \sin\varphi_{kj} \right) \\
D_{kij} &= \sigma_{ki}\sigma_{kj}\sqrt{\alpha_{ki}\alpha_{kj}}
\left( \sin\varphi_{ki}\sin\varphi_{kj}
                                    + \cos\varphi_{ki} \cos\varphi_{kj}
                                    \right).
\end{aligned}
\end{equation}
In order to satisfy the second set of conditions, expressed in
Eqs.~\eqref{eqn:rotation-matrix-constraintsB}, one finds that, for
all~$i,j$ and~$k$, either~$\sigma_{ki} = \sigma_{kj}$
or~$\alpha_{ki} \alpha_{kj} = 0$ must hold.  Hence, when written
in the form in Eq.~\eqref{eqn:restricted-rotation}, the non-zero
$\rmatrix{T}$~sub-matrices in a given row of~$\rmatrix{M}$ are
either all scale-rotation or all scale-reflection-rotation
matrices.

Since~$\rmatrix{M}$ represents the mapping,~$\map{M}$, and, by
Postulate~3.1,~$\map{M}^{-1}$ exists, the
matrix~$\rmatrix{M}^{-1}$ represents the mapping~$\map{M}^{-1}$.
Hence, the matrix~$\rmatrix{M}^{-1} = \rmatrix{M}^T$, must also
satisfy Postulate~3.2.  Now, from
Eq.~\eqref{eqn:restricted-rotation}, the matrix~$\rmatrix{M}^T$
takes the form
\begin{equation}
\rmatrix{M}^T =
                \begin{pmatrix}
                \label{eqn: restricted-rotation}
                \left(T^{(11)}\right)^T &  \left(T^{(21)}\right)^T & \dots &
                \left(T^{(N1)}\right)^T \\
                \left(T^{(12)}\right)^T &  \left(T^{(22)}\right)^T & \dots &
                \left(T^{(N2)}\right)^T \\
                    \hdotsfor[2]{4}                 \\
                 \left(T^{(1N)}\right)^T &  \left(T^{(2N)}\right)^T & \dots &
                \left(T^{(NN)}\right)^T.
\end{pmatrix}
\end{equation}
In order to satisfy Postulate~3.2, the non-zero sub-matrices
of~$\rmatrix{M}^T$ in a given row are either all scale-rotation or
all scale-reflection-rotation matrices.  But this implies that,
in~$\rmatrix{M}$, the non-zero $\rmatrix{T}$~sub-matrices in a
given \emph{column} are either all scale-rotation or all
scale-reflection-rotation matrices. Hence,  the non-zero
$\rmatrix{T}$~sub-matrices that compose the matrix~$\rmatrix{M}$
are either \emph{all} scale-rotation or \emph{all}
scale-reflection-rotation matrices.

\medskip
\paragraph{Recasting~$\rmatrix{M}$ as a complex transformation}
\label{para:R-recast-as-unitary}

At this point, it is convenient to recast the effect of
$\rmatrix{M}$ on the state in a complex form. Let the complex form
of the state,~$\lvect{Q}$, be defined as
\begin{equation}  \label{eqn:complex-form-of-state}
    \cvect{v}=
    \begin{pmatrix}
        Q_1 + iQ_2 \\
        Q_3 + iQ_4 \\
        \dots \\
        Q_{2N-1}  + iQ_{2N}
    \end{pmatrix},
\end{equation}
and let us suppose that the~$\cvect{v}$ are vectors in a complex
vector space with inner product,~$\langle \cvect{u}, \cvect{v}
\rangle= \sum_i u_i^* v_i$ and norm~$|\cvect{v}|=\sqrt{\langle
\cvect{v}, \cvect{v} \rangle}$. Consider the action of the
$N$-dimensional complex matrix,~$\cmatrix{V}$, on~$\cvect{v}$,
\begin{equation} \label{eqn:complex-transformationA}
\cvect{v}' = \cmatrix{V} \cvect{v},
\end{equation}
where~$\cvect{v}'$ is defined analogously to~$\cvect{v}$. By
multiplying out the real and complex parts of this expression, it
can be seen that the effect of~$\cmatrix{V}$ on~$\cvect{v}$ is
equivalent to the action of the real $2N$-dimensional
matrix,~$\rmatrix{M}_V$, on~$\lvect{Q}$,
\begin{equation} \lvect{Q}' = \rmatrix{M}_V \lvect{Q},
\end{equation}
with
\begin{equation}
    M_V =
       \begin{pmatrix}
        \cmatrix{V}_{11}^R & -\cmatrix{V}_{11}^I  & \dots  & \dots & \cmatrix{V}_{1N}^R &
                                                                -\cmatrix{V}_{1N}^I  \\
        \cmatrix{V}_{11}^I & \cmatrix{V}_{11}^R & \dots  & \dots &  \cmatrix{V}_{1N}^I &
                                                                \cmatrix{V}_{1N}^R  \\
        \hdotsfor{6} \\
        \hdotsfor{6} \\
        \cmatrix{V}_{N1}^R & -\cmatrix{V}_{N1}^I & \dots  & \dots & \cmatrix{V}_{NN}^R &
                                                                -\cmatrix{V}_{NN}^I  \\
        \cmatrix{V}_{N1}^I & \cmatrix{V}_{N1}^R & \dots  & \dots & \cmatrix{V}_{NN}^I &
                                                                \cmatrix{V}_{NN}^R
    \end{pmatrix},
\end{equation}
where~$\cmatrix{V}_{ij}^R$ and~$\cmatrix{V}_{ij}^I$ are,
respectively, the real and imaginary parts of~$\cmatrix{V}_{ij}$.
If~$\cmatrix{V}_{ij}$ is chosen to be $\sqrt{\alpha_{ij}}
\exp{i\varphi_{ij}}$, then~$M_V$ becomes identical to~$M$ in the
case where the non-zero $\rmatrix{T}$~sub-matrices
of~$\rmatrix{M}$ consist of scale-rotations.

The orthogonality of~$M_V$ implies that~$\cmatrix{V}$ is unitary.
 To see this, consider
\begin{equation} \label{eqn:VVij}
    (\cmatrix{V}^\dag \cmatrix{V})_{ij} =
            \sum_k \sqrt{\alpha_{ki}\alpha_{kj}} e^{i(\varphi_{kj} - \varphi_{ki})}.
\end{equation}
Denote by~$\lvect{M}_q$ the $2N$-dimensional real vector formed
from the~$q$th column of~$M_V$, and let the relations in
Eqs.~\eqref{eqn:matrix-definitionsB}
and~\eqref{eqn:matrix-definitionsA} be defined
for~$\rmatrix{M}_V$.  Then, from
Eqs.~\eqref{eqn:matrix-definitionsA}, $(\cmatrix{V}^\dag
\cmatrix{V})_{ii} = \sum_k \alpha_{ki}$ is $|\lvect{M}_{2i-1}|^2$,
which is unity since~$M_V$ is an orthogonal matrix.  To
evaluate~$(\cmatrix{V}^\dag \cmatrix{V})_{ij}$ for~$i \neq j$, it
is helpful to rewrite~$A_{kij}$ and~$B_{kij}$ in terms
of~$\cmatrix{V}_{ij}$,
\begin{align}
        A_{kij} &= \cmatrix{V}_{ki}^R \cmatrix{V}_{kj}^R + \cmatrix{V}_{ki}^I \cmatrix{V}_{kj}^I
         \\
        -B_{kij} &= \cmatrix{V}_{ki}^R \cmatrix{V}_{kj}^I - \cmatrix{V}_{ki}^I \cmatrix{V}_{kj}^R
\end{align}
so that
\begin{equation}
    \begin{split}
     \cmatrix{V}_{ki}^*\cmatrix{V}_{kj}
     &= (\cmatrix{V}_{ki}^R \cmatrix{V}_{kj}^R + \cmatrix{V}_{ki}^I \cmatrix{V}_{kj}^I)
                +i(\cmatrix{V}_{ki}^R \cmatrix{V}_{kj}^I - \cmatrix{V}_{ki}^I
                \cmatrix{V}_{kj}^R) \\
     &= A_{kij} - iB_{kij}
     \end{split}
\end{equation}
and
\begin{equation}
    \begin{split}
     \sum_{k=1}^{N} \cmatrix{V}_{ki}^* \cmatrix{V}_{kj}
     &= \sum_{k=1}^{N} A_{kij} - iB_{kij} \\
     &= \lvect{M}_{2i-1} \cdot \lvect{M}_{2j-1}
      -i \lvect{M}_{2i-1} \cdot \lvect{M}_{2j},
    \end{split}
\end{equation}
which, due to the orthogonality of~$\rmatrix{M}$, is zero
whenever~$i \neq j$.  Therefore, $(\cmatrix{V}^\dag
\cmatrix{V})_{ij} = \delta_{ij}$, so that~$\cmatrix{V}$ is
unitary.

Similarly, if one considers the effect of the complex
transformation~$\cmatrix{V}\cmatrix{K}$, where~$\cmatrix{K}$ is
the complex conjugation operation, acting on~$\cvect{v}$,
\begin{equation} \label{eqn:complex-transformationB}
\cvect{v}' = \cmatrix{V}\cmatrix{K} \cvect{v},
\end{equation}
one finds that this is equivalent to the action of the matrix~$M$
on~$\lvect{Q}$ in the case that the non-zero
$\rmatrix{T}$~sub-matrices that comprise~$\rmatrix{M}$ are
scale-reflection-rotation matrices. Since~$\cmatrix{V}$ is
unitary, the transformation~$\cmatrix{V}\cmatrix{K}$ is
antiunitary.

Thus far, we have shown only that the complex
transformations~$\cmatrix{V}$ and~$\cmatrix{V}\cmatrix{K}$ satisfy
Postulate~3.2 in the special cases of~$\lvect{Q}$ examined above.
To show that these transformations satisfy Postulate~3.2 for any
state, note that the addition of~$\var_0$ to each of the~$\var_i$
in the complex form of the state,~$\cvect{v}$, generates the
vector~$e^{i\var_0} \cvect{v}$, that is
\begin{equation} \label{eqn:invariance-of-P'-A}
\cvect{v} \xrightarrow{+ \var_0} e^{i\var_0} \cvect{v}.
\end{equation}
As a result, the vector~$\cvect{v}'$ in
Eq.~\eqref{eqn:complex-transformationA} transforms as
\begin{equation} \label{eqn:invariance-of-P'-B}
\cvect{v}' \xrightarrow{+ \var_0} e^{i\var_0} \cvect{v}',
\end{equation}
and the vector~$\cvect{v}'$ in
Eq.~\eqref{eqn:complex-transformationB} transforms as
\begin{equation} \label{eqn:invariance-of-P'-C}
\cvect{v}' \xrightarrow{+ \var_0} e^{-i\var_0} \cvect{v}',
\end{equation}
Since the~$P_i'$ are independent of the overall phase
of~$\cvect{v}'$, it follows that, in both
Eqs.~\eqref{eqn:invariance-of-P'-B}
and~\eqref{eqn:invariance-of-P'-C}, the~$P_i'$ remain unchanged by
the addition of~$\var_0$ to the~$\var_i$. Therefore, the
transformations~$\cmatrix{V}$ and~$\cmatrix{V}\cmatrix{K}$ both
satisfy Postulate~3.2.

\subsubsection{Step 3: General Unitary and Antiunitary Transformations}
\label{sec:every-unitary-is-ok}

We have shown so far that the imposition of Postulate~3.2
restricts~$\rmatrix{M}$ to a subset of the set of orthogonal
transformations, and that each transformation in this subset can
be recast as either a unitary or an antiunitary transformation.
But, we have not ruled out the possibility that there are unitary
or antiunitary transformations which are not equivalent to
orthogonal transformations satisfying Postulate~3.2.  In this
section, it shall be shown that, in fact, \emph{any}
$N$-dimensional unitary or antiunitary transformation satisfies
Postulates~3.1,~3.2, and~4.

Consider the arbitrary unitary transformation~$\cmatrix{U}$.  The
transformation
\begin{equation} \label{eqn:general-U-on-v}
\cvect{v}' = \cmatrix{U} \cvect{v}
\end{equation}
is equivalent to the transformation
\begin{equation} \lvect{Q}' =
\rmatrix{M} \lvect{Q},
\end{equation}
where
\begin{equation} \label{eqn:matrix-M-U}
\rmatrix{M} =
\begin{pmatrix}
        \cmatrix{U}_{11}^R    & -\sigma\cmatrix{U}_{11}^I    & \dots  & \dots  & \cmatrix{U}_{1N}^R  & -\sigma\cmatrix{U}_{1N}^I  \\
        \cmatrix{U}_{11}^I    & \sigma\cmatrix{U}_{11}^R     & \dots  & \dots  & \cmatrix{U}_{1N}^I  & \sigma\cmatrix{U}_{1N}^R  \\
        \hdotsfor{6} \\
        \hdotsfor{6} \\
        \cmatrix{U}_{N1}^R & -\sigma\cmatrix{U}_{N1}^I     & \dots     & \dots     & \cmatrix{U}_{NN}^R &  -\sigma\cmatrix{U}_{NN}^I  \\
        \cmatrix{U}_{N1}^I & \sigma\cmatrix{U}_{N1}^R     & \dots     & \dots     & \cmatrix{U}_{NN}^I &   \sigma\cmatrix{U}_{NN}^R
    \end{pmatrix},
\end{equation}
with~$\sigma=1$.  Similarly, using the arbitrary antiunitary
transformation~$\cmatrix{U}\cmatrix{K}$, one finds the
corresponding matrix to be~$\rmatrix{M}$ with~$\sigma=-1$.

First we show that~$M$ is an orthogonal matrix. In the
following,~$\lvect{M}_q$ denotes the real $2N$-dimensional vector
formed from the~$q$th column of~$M$.

$M$ is an orthogonal matrix since:
\begin{description}
    \item[(a)] the columns of~$M$ are normalized:
    \begin{equation}
    \begin{split}
        |\lvect{M}_{2i-1}|^2    &= |\lvect{M}_{2i}|^2 \qquad \text{from Eq.~\eqref{eqn:matrix-M-U}}\\
                                &= \sum_{k=1}^{N} |\cmatrix{U}_{ki}|^2        \\
                                &= 1                        \qquad\qquad \text{since~$\cmatrix{U}$ is
                                unitary}
    \end{split}
    \end{equation}

    \item[(b)] the columns of~$M$ are orthogonal:

    \begin{description}
        \item[(i)] Columns~$(2i-1)$ and $2i$, for~$i=1, 2, \dots, N$,
        are orthogonal since, from Eq.~\eqref{eqn:matrix-M-U},
        \begin{equation} \label{eqn:R-U-columns1}
            \lvect{M}_{2i-1} \cdot \lvect{M}_{2i} =0.
        \end{equation}

        \item[(ii)] By inspection of Eq.~\eqref{eqn:matrix-M-U}, one sees
        that, for~$i \neq j$,
        \begin{equation}
            \begin{gathered}
                \lvect{M}_{2i-1} \cdot \lvect{M}_{2j-1}   = \lvect{M}_{2i} \cdot \lvect{M}_{2j}\\
                \lvect{M}_{2i-1} \cdot \lvect{M}_{2j} =-\lvect{M}_{2i} \cdot \lvect{M}_{2j-1}.
            \end{gathered}
        \end{equation}

        But, since~$\cmatrix{U}$ is unitary,
        \begin{equation}
            \begin{split}
            \sum_{k=1}^{N} \cmatrix{U}_{ki}^* \cmatrix{U}_{kj}
                &= \lvect{M}_{2i-1} \cdot \lvect{M}_{2j-1}
                            -i\lvect{M}_{2i-1} \cdot \lvect{M}_{2j} \\
                &= 0, \qquad i \neq j.
            \end{split}
        \end{equation}

        Therefore, for~$i \neq j$,
        \begin{equation}\label{eqn:R-U-columns2}
            \begin{gathered}
                \lvect{M}_{2i-1} \cdot \lvect{M}_{2j-1}
                        = \lvect{M}_{2i} \cdot \lvect{M}_{2j} =0\\
                 \lvect{M}_{2i-1}\cdot \lvect{M}_{2j}
                        =-\lvect{M}_{2i} \cdot \lvect{M}_{2j-1}=0.
            \end{gathered}
        \end{equation}
    \end{description}


\end{description}

Since~$\rmatrix{M}$ is an orthogonal matrix, it satisfies
Postulates~3.1 and~4. The invariance of the~$P_i'$ required by
Postulate~3.2 follows from the observation that, under the
addition of~$\var_0$ to the~$\var_i$ in~$\cvect{v}$,
\begin{equation} \label{eqn:invariance-of-P'-A'}
\cvect{v} \xrightarrow{+ \var_0} e^{i\var_0} \cvect{v}.
\end{equation}
As a result, the vector~$\cvect{v}'$ in
Eq.~\eqref{eqn:general-U-on-v} transforms as
\begin{equation} \label{eqn:invariance-of-P'-B'}
\cvect{v}' \xrightarrow{+ \var_0} e^{i\sigma \var_0} \cvect{v}',
\end{equation}
with~$\sigma= \pm 1$ depending upon whether a unitary or antiunitary
transformation is chosen. In either case, since the~$P_i'$ are
independent of the overall phase of~$\cvect{v}'$, it follows that
the~$P_i'$ remain invariant.

Hence, any unitary or antiunitary transformation satisfies
Postulates~3.1,~3.2, and~4.

\subsubsection{Step 4: Physical Transformations}

By Postulate~3.3, a physical transformation~(such as a
reflection-rotation of a frame of reference) that depends
continuously upon a real-valued parameter n-tuple~$\Bpi$ is
represented by a map~$\map{M}_{\Bpi}$ which depends continuously
upon~$\Bpi$.  From Eq.~\eqref{eqn:matrix-M-U}, the
matrix~$\rmatrix{M}_{\Bpi}$, which represents~$\map{M}_{\Bpi}$,
contains the discrete parameter~$\sigma$. Given two
$\rmatrix{M}$-matrices,~$\rmatrix{M}$ and~$\rmatrix{M}'$, with
\emph{different} values of~$\sigma$, it follows from
Eq.~\eqref{eqn:matrix-M-U} that it is only possible to continuously
transform~$\rmatrix{M}$ into~$\rmatrix{M}'$ provided
that~$\rmatrix{M}$ can pass through the null matrix.
However,~$\rmatrix{M}$ cannot be null since this would require that
the~$U_{ij}$ simultaneously vanish, which is impossible
since~$\cmatrix{U}$ is unitary. Therefore, it is not possible to
continuously transform between two $\rmatrix{M}$-matrices with
different values of~$\sigma$. Hence, the matrix~$\rmatrix{M}_{\Bpi}$
has~$\sigma=1$ or~$\sigma=-1$ for all~$\Bpi$, which implies that the
physical transformation under discussion is represented either by
unitary~($\sigma=1$) or antiunitary~($\sigma=-1$) transformations.

Furthermore, by Postulate~3.3, a \emph{continuous} physical
transformation that depends continuously upon a real-valued
parameter n-tuple~$\Bpi$ is represented by a map~$\map{M}_{\Bpi}$
which reduces to the identity map for some value of~$\Bpi$.  From
Eq.~\eqref{eqn:matrix-M-U}, we see that, for~$\sigma =1$, the
matrix~$\rmatrix{M}$ consists of scale-rotation sub-matrices
which, with a suitable choice of the~$\alpha_{ij}$ and
the~$\varphi_{ij}$, reduces to the identity. However,
with~$\sigma=-1$, it can be seen that a reduction to the identity
is not possible. 
Therefore, a continuous physical transformation can only be
represented by unitary transformations~($\sigma=1$).

Finally, a discrete physical transformation~(such as temporal
inversion) is represented by a matrix~$\rmatrix{M}$ in which
either~$\sigma =1$ or~$\sigma =-1$, and is therefore represented by
either a unitary or an antiunitary transformation.

\subsection{Representation of Measurements} \label{sec:D4}

In the previous section, it has been shown that the state of a
system at time~$t$ that has been prepared by a measurement
in~$\set{A}$ can, from the point of view of a measurement~$\mment{A}
\in \set{A}$, be represented as the complex vector
\begin{equation}
\cvect{v} = \begin{pmatrix}
            \sqrt{P_1} e^{i\var_1} \\
            \sqrt{P_2} e^{i\var_2} \\
            \ldots                 \\
            \sqrt{P_N} e^{i\var_N} \\
            \end{pmatrix},
\end{equation}
where the~$P_i$ are the outcome probabilities of
measurement~$\mment{A}$ if performed at time~$t$. Furthermore, it
has been shown that any interaction following the preparation can
be represented by a unitary transformation of~$\cvect{v}$.

Consider an experiment where a system undergoes some
measurement~$\mment{A} \in \set{A}$, yields a particular outcome,
and subsequently undergoes some other measurement~$\mment{A}' \in
\set{A}$ that may or may not be the same as~$\mment{A}$.  The
purpose of this section is to develop the formalism necessary to
predict the outcome probabilities in such an experiment.

\subsubsection{Prepared States}

Suppose that, in the above-mentioned experiment, a system
undergoes measurement~$\mment{A}$ and yields outcome~$j$. What is
the state of the prepared system?

By Postulate~1.1, measurement~$\mment{A}$ has~$N$ possible outcomes
and, by the assumption of repetition
consistency~(Sec.~\ref{sec:F-idealised-set-up}), after~$\mment{A}$
has been performed and outcome~$j$ obtained, immediate repetition
yields the same outcome with certainty. Therefore, for every
outcome~$j$ there exists a corresponding state,~$\cvect{v}_j$, such
that the measurement~$\mment{A}$ upon the system in
state~$\cvect{v}_j$ yields outcome~$j$ with certainty. From
Eq.~\eqref{eqn:complex-form-of-state}, since~$P_j = 1$ and all the
other~$P_j$ are zero, we have that
\begin{equation} \label{eqn:form-of-v-j}
\cvect{v}_j = ( 0,\dots, e^{i\var_j}, \dots, 0)^{\text{\textsf{T}}},
\end{equation}
where~$\var_j$ is undetermined.

\subsubsection{Measurements}

\begin{figure}[!h]
\begin{centering}
\includegraphics[width=3.25in]{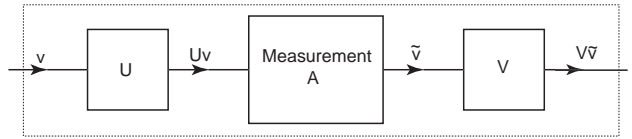}
\caption{\label{fig:representation-of-measurement} A
representation of a measurement of~$\mment{A}'$.  A unitary
transformation,~$\cmatrix{U}$, transforms the input
state,~$\cvect{v}$, into~$\cmatrix{U}\cvect{v}$.
Measurement~$\mment{A}$ is performed on this state, and the output
state,~$\tilde{\cvect{v}}$, of the measurement is transformed by
the unitary transformation~$\cmatrix{V}$
into~$\cmatrix{V}\tilde{\cvect{v}}$.}
\end{centering}
\end{figure}

By Postulate~1.2, measurement~$\mment{A}'$ can be represented by
an arrangement consisting of a measurement~$\mment{A}$ followed
immediately before and after by suitable interactions. These
interactions bring about continuous transformations of the system.
From the results of the previous section, these interactions must,
therefore, be represented by unitary transformations, which we
shall denote~$\cmatrix{U}$ and~$\cmatrix{V}$, respectively~(see
Fig.~\ref{fig:representation-of-measurement}).  In the following,
we shall establish the form of these matrices, and then obtain an
expression for the outcome probabilities for
measurement~$\mment{A}'$ performed on a system in
state~$\cvect{v}$.

First, from Postulate~1.1 and the assumption of repetition
consistency, there exist~$N$ states~$\cvect{v}_1', \cvect{v}_2',
\ldots, \cvect{v}_N'$ such that measurement~$\mment{A}'$ performed
on a system in state~$\cvect{v}_i'$ yields outcome~$i$ with
certainty. Hence, the arrangement in
Fig.~\ref{fig:representation-of-measurement} must be such
that~$\mment{A}$ yields outcome~$i$ with certainty when the input
state to the arrangement is~$\cvect{v}_i'$.  For this to be the
case,~$\cmatrix{U}$ must transform~$\cvect{v}_i'$ to a state of
the form~$\cvect{v}_i e^{i\vard_i}$, where~$\vard_i$ is arbitrary.
That is, the matrix~$\cmatrix{U}$ must satisfy the relations
\begin{equation} \label{eqn:measurement-relations-for-U}
    \cmatrix{U} \cvect{v}_i' = \cvect{v}_i e^{i\vard_i},
    \quad i =1,2, \dots, N
\end{equation}

Second, if outcome~$i$ is obtained from the arrangement, the
output state of the arrangement must be of the form~$\cvect{v}_i'
e^{i\vard_i'}$, where~$\vard_i'$ is arbitrary.  But, immediately
after measurement~$\mment{A}$, the system is in
state~$\cvect{v}_i$ up to an overall phase.  Hence, the
matrix~$\cmatrix{V}$ must satisfy the relations
\begin{equation} \label{eqn:measurement-relations-for-V}
    \cmatrix{V} \cvect{v}_i = \cvect{v}_i' e^{i \vard_i'}
    \quad i =1,2, \dots, N
\end{equation}

From Eq.~\eqref{eqn:form-of-v-j}, the~$\cvect{v}_i$ form an
orthonormal basis for~$\numberfield{C}^N$, and, from
Eq.~\eqref{eqn:measurement-relations-for-U},~$\cvect{v}_i' =
\cmatrix{U}^\dag \cvect{v}_i e^{i\vard_i}$, which,
since~$\cmatrix{U}$ is unitary, implies that the~$\cvect{v}_i'$ also
form an orthonormal basis.  Therefore, any state,~$\cvect{v}$, can
be expanded as~$\sum_i c_i' \cvect{v}_i'$, with~$c_i' \in
\numberfield{C}$, and the matrices~$\cmatrix{U}$ and~$\cmatrix{V}$
are determined by the relations in
Eqs.~\eqref{eqn:measurement-relations-for-U}
and~\eqref{eqn:measurement-relations-for-V} up to the~$\vard_i$ and
the~$\vard_i'$.

It is now possible to determine the outcome probabilities if a
system in state~$\cvect{v}$ undergoes measurement~$\mment{A}'$.
Using Eq.~\eqref{eqn:measurement-relations-for-U} and the
expansion~$\cvect{v} = \sum_i c_i' \cvect{v}_i'$, the first
interaction of the arrangement transforms~$\cvect{v}$ into
\begin{equation}
\cmatrix{U}\left(\sum_i c_i'\cvect{v}_i'\right) = \sum_i
c_i'\cvect{v}_i e^{i\vard_i}.
\end{equation}
The probability that measurement~$\mment{A}$ in the arrangement
yields outcome~$i$ is therefore~$|c_i'|^2$. Hence,
measurement~$\mment{A}'$ performed on the state~$\cvect{v}$ yields
outcome~$i$ with probability~$|c_i'|^2$.


In summary, every measurement,~$\mment{A}' \in \set{A}$, has an
associated orthonormal basis,~$\{ \cvect{v}_1', \cvect{v}_2',
\dots, \cvect{v}_N' \}$.  Such a measurement can be implemented by
a measurement~$\mment{A}$ followed immediately before and after by
interactions represented by~$\cmatrix{U}$ and~$\cmatrix{V}$
defined in Eqs.~\eqref{eqn:measurement-relations-for-U}
and~\eqref{eqn:measurement-relations-for-V} in terms of these
basis vectors. If measurement~$\mment{A}'$ is performed upon a
system in state~$\cvect{v}$, the probability,~$P_i'$, of obtaining
outcome~$i$ is~$|c_i'|^2$, where~$c_i'$ is determined by the
relation~$\cvect{v} = \sum_i c_i' \cvect{v}_i'$.

\subsubsection{Expected Values}

If the $i$th~outcome of measurement~$\mment{A}'$ has an associated
real value~$a_i'$, the expected value obtained in an experiment in
which a system in state~$\cvect{v}$ undergoes
measurement~$\mment{A}'$ is defined as
\begin{equation}
\langle \cmatrix{A}' \rangle = \sum_i a_i' P_i'.
\end{equation}
Since~$P_i' = |c_i'|^2$ and~$c_i' = \cvect{v}_i'^\dag \cvect{v}$,
this expression can be also written as
\begin{equation}
\begin{split}
\langle \cmatrix{A}' \rangle &= \sum_i \cvect{v}^\dag
                    \left(\cvect{v}_i'
                        a_i'\cvect{v}_i'^\dag \right) \cvect{v} \\
                    &= \cvect{v}^\dag \left(\sum_i \cvect{v}_i'
                        a_i'\cvect{v}_i'^\dag \right) \cvect{v} \\
                    &= \cvect{v}^\dag \cmatrix{A}' \cvect{v},
\end{split}
\end{equation}
where the matrix~$\cmatrix{A}' \equiv \sum_i \cvect{v}_i'
a_i'\cvect{v}_i'^\dag$ is Hermitian since the~$a_i'$ are real, and
is non-degenerate since the~$a_i'$ have been assumed to be
distinct~(Sec.~\ref{sec:F-idealised-set-up}).

Since the~$\cvect{v}_i'$ are eigenvectors of~$\cmatrix{A}'$, with
the~$a_i'$ being the corresponding eigenvalues, the
matrix~$\cmatrix{A}'$ provides a compact mathematical way of
representing all the relevant details about
measurement~$\mment{A}'$.

\subsection{Composite Systems}
\label{sec:D5}

 It is often the case that a given physical system
can be subject to examination in distinct experimental set-ups,
where, loosely speaking, the measurements in each set-up probe
distinct properties of the system.  Formally, we can express this
as follows.

Consider a system which admits abstract quantum
model,~$\model{q}(N^{(1)})$, with respect to measurement
set~$\set{A}^{(1)}$, and which admits abstract quantum
model,~$\model{q}(N^{(2)})$, with respect to measurement
set~$\set{A}^{(2)}$, where the set-ups defined by measurement
sets~$\set{A}^{(1)}$ and~$\set{A}^{(2)}$ are disjoint~(in the sense
defined in Sec.~\ref{sec:F}). The system can also be modeled as a
whole. That is, we can construct the measurement set~$\set{A} =
\set{A}^{(1)} \times \set{A}^{(2)}$, and construct abstract quantum
model~$\model{q}(N)$, where~$N = N^{(1)}N^{(2)}$. We shall
accordingly speak of the system as a \emph{composite} system
consisting of two \emph{sub-systems}. More generally, if a system
admits~$d$~($d>1$) abstract quantum models with respect to
$d$~disjoint measurement sets, we shall speak of it as a composite
system consisting of $d$~sub-systems.

One often prepares a state of a composite system by first preparing
each of its subsystems, and then allowing these subsystems to
interact with one another.  In order to formally describe such a
procedure, one needs a rule, the composite system rule, which we
shall now derive, that enables the state of the system to be written
down in terms of the states of its sub-systems.

\subsubsection*{The Composite System Rule}

In order to derive the composite system rule, we shall apply
Postulate~5 to the case of a composite system with two sub-systems
with abstract models~$\model{q}(N^{(1)})$
and~$\model{q}(N^{(2)})$, respectively, where the composite system
has the abstract model~$\model{q}(N)$.

Suppose that the sub-systems are in states represented
as~$(P_i^{(1)}; \var_i^{(1)})$ and~$(P_j^{(2)}; \var_j^{(2)})$,
respectively. Then, by Postulate~5, the state of the composite
system can be represented as~$(P_{ij}; \var_{ij})$, where
\begin{gather} \label{eqn:sub-system-postulate1}
P_{ij} = P_i^{(1)} P_j^{(2)} \\
\label{eqn:sub-system-postulate2} \var_{ij} = \var_i^{(1)} +
\var_j^{(2)}.
\end{gather}
If we write the states of the sub-systems in complex form,
        \begin{equation*}\cvect{v}^{(1)} = \left(
                             \sqrt{P_1^{(1)}}e^{i\var_1^{(1)}},
                             \sqrt{P_2^{(1)}}e^{i\var_2^{(1)}},
                             \dots,
                             \sqrt{P_{N^{(1)}}^{(1)}}e^{i\var_{N^{(1)}}^{(1)}}\right)^{\text{\textsf{T}}}
        \end{equation*}
and
        \begin{equation*}
        \cvect{v}^{(2)} = \left(
                            \sqrt{P_1^{(2)}}e^{i\var_1^{(2)}},
                            \sqrt{P_2^{(2)}}e^{i\var_2^{(2)}},
                            \cdots,
                            \sqrt{P_{N^{(2)}}^{(1)}}e^{i\var_{N^{(2)}}^{(2)}}\right)^{\text{\textsf{T}}},
        \end{equation*}
respectively, and, similarly, write the state of the composite
system as
        \begin{equation*}
        \cvect{v} = \left(
                             \sqrt{P_{11}}e^{i\var_{11}},
                             \cdots,
                             \sqrt{P_{N^{(1)}, N^{(2)}}} e^{i\var_{N^{(1)},
                             N^{(2)}}}\right)^{\text{\textsf{T}}},
        \end{equation*}
then it follows from Eqs.~\eqref{eqn:sub-system-postulate1}
and~\eqref{eqn:sub-system-postulate2}
 that~$\cvect{v}$ can simply be
written as~$\cvect{v}^{(1)} \otimes \cvect{v}^{(2)}$.

More generally, consider a composite system with $d$~sub-systems,
numbered~$1, 2, \dots, d$, in states~$\cvect{v}^{(1)},
\cvect{v}^{(2)}, \dots, \cvect{v}^{(d)}$, respectively.  We can
regard sub-systems~$1$ and~$2$ as comprising a bipartite composite
system, system~$1'$, which, according to the above result, is in
state~$\cvect{v}^{(1)} \otimes \cvect{v}^{(2)}$. Next, we can
regard system~$1'$ and sub-system~$3$ as comprising a bipartite
composite system, system~$2'$, which is therefore in
state~$(\cvect{v}^{(1)} \otimes \cvect{v}^{(2)}) \otimes
\cvect{v}^{(3)}$.  Continuing in this way, we can see the state of
the composite system with $d$~sub-systems has the state~$\cvect{v}
= \cvect{v}^{(1)} \otimes \cvect{v}^{(2)} \otimes \dots \otimes
\cvect{v}^{(d)}$.

\subsection{Some Generalizations}

\subsubsection{Representation of sub-system measurements}

Suppose that measurement~$\mment{A}^{(1)} \in \set{A}^{(1)}$,
represented by $N^{(1)}$-dimensional Hermitian
operator~$\cmatrix{A}^{(1)}$, with eigenstates~$\cvect{v}_i^{(1)}$
and eigenvalues~$a_i$, respectively, is performed on sub-system~1
of a bipartite composite system. With respect to the abstract
quantum model~$\model{q}(N)$ of the composite system,
measurement~$\mment{A}^{(1)}$ is \emph{not} in the measurement
set~$\set{A}$ of the composite system since the measurement has
only~$N^{(1)}$ distinct outcomes whereas a measurement
in~$\set{A}$ has~$N = N^{(1)} N^{(2)} > N^{(1)}$ possible
outcomes.  However, it is convenient to be able to describe
measurement~$\mment{A}^{(1)}$, which we shall describe as a
\emph{sub-system measurement}, as an $N$-dimensional
operator~$\cmatrix{A}$, in the framework of~$\model{q}(N)$.

To determine the form of~$\cmatrix{A}$, it is sufficient to consider
the effect of~$\cmatrix{A}$ on product states of the
form~$\cvect{v}_i^{(1)} \otimes \cvect{v}^{(2)}$ of the composite
system, where~$\cmatrix{A}^{(1)}\cvect{v}_i^{(1)} = a_i
\cvect{v}_i^{(1)}$.  If the composite system is in such a state,
then sub-system~1 is in state~$\cvect{v}_i^{(1)}$. Therefore, when
measurement~$\mment{A}^{(1)}$ is performed, outcome~$a_i$ is
obtained with certainty, and the state of sub-system~1 is
unchanged~(up to an irrelevant overall phase). Therefore, the state
of the composite system remains unchanged.  If we require
that~$\cmatrix{A}$ has eigenvectors~$\cvect{v}_i^{(1)} \otimes
\cvect{v}^{(2)}$, with respective eigenvalues~$a_i$, it follows
that~$\cmatrix{A}$ can be taken to be~$\cmatrix{A}^{(1)} \otimes
\cmatrix{I}^{(2)}$, where~$\cmatrix{I}^{(2)}$ is the identity matrix
in the model of sub-system~2, with the only freedom being a
physically irrelevant overall phase in each of the eigenstates
of~$\cmatrix{A}^{(1)}$.

The above result trivially generalizes to the case of a measurement
performed on one sub-system of a composite system consisting of
$d$~sub-systems.

\subsubsection{Degenerate measurements}

The model~$\model{q}(N)$, whose explicit mathematical form has been
derived above, applies to an abstract set-up where the measurements,
chosen from the set~$\set{A}$, have~$N$ possible outcomes and
therefore, by the distinctness assumption of
Sec.~\ref{sec:F-idealised-set-up}, necessarily have~$N$ distinct
outcome values. From the above discussions, it follows that each
measurement~$\mment{A} \in \set{A}$ is represented by a
non-degenerate Hermitian operator of dimension~$N$.

Now, it is useful to be able to describe measurements within the
context of model~$\model{q}(N)$ which have fewer than~$N$ outcomes.
An example of such measurements that we have discussed above are
sub-system measurements. We shall now broaden the discussion to
allow for measurements with~$N'<N$ possible outcomes where~$N'$ is
not a multiple of~$N$ and which therefore cannot be regarded as
sub-system measurements.

Consider an abstract set-up where a preparation implemented using a
measurement from~$\set{A}$ is followed by measurement~$\mment{A}$,
whose observable outcome probabilities are denoted~$P_1, \dots,
P_N$. Suppose that, if measurement~$\mment{B}$~(with~$N'<N$)
possible outcomes) replaces measurement~$\mment{A}$, the outcome
probabilities,~$P_1', \dots, P_{N'}'$ of measurement~$\mment{B}$ can
be determined from the~$P_i$ by a many-to-one map of the outcomes
of~$\mment{A}$ to the outcomes of~$\mment{B}$.  For example, in the
case where~$N=3$ and~$N'=2$, the map from the outcomes
of~$\mment{A}$ to the outcomes of~$\mment{B}$ might consist in~$1
\rightarrow 1'$, $2 \rightarrow 2'$ and~$3 \rightarrow 2'$, in which
case~$P_1' = P_1$ and~$P_2' = P_2 + P_3$.   In such a case, we shall
say that measurement~$\mment{B}$ is a \emph{degenerate form} of
measurement~$\mment{A}$; or, more simply, that
measurement~$\mment{B}$ is a degenerate measurement.

Now, measurement~$\mment{B}$ can \emph{formally} be treated as if it
has~$N$ possible outcomes, but where some of these outcomes have the
\emph{same} value.  In this mode of description, in the above
example, one can maintain a one-to-one map between the outcomes
of~$\mment{A}$ and of~$\mment{B}$~(so that~$1 \rightarrow 1'$, $2
\rightarrow 2'$ and so on), but label the outcomes of~$\mment{B}$
with their outcome values, and, when computing the outcome
probabilities of~$\mment{B}$, group together the outcomes with the
same outcome value.  In the above example, one would respectively
label the three outcomes with outcome values~$b_1, b_2$ and~$b_3$,
and but have~$b_2=b_3$.

Since measurement~$\mment{B}$ is a degenerate form of
measurement~$\mment{A}$, it can be represented by the
$N$-dimensional degenerate Hermitian operator~$\cmatrix{B} = \sum_i
b_i \cvect{v}_i \cvect{v}_i^\dagger$, where~$\cmatrix{A}\cvect{v}_i
= a_i \cvect{v}_i$.  The outcome probabilities for
measurement~$\mment{B}$ can then be computed in the usual way, on
the understanding that those outcomes with the same outcome values
must not be regarded as physically distinguishable, but must be
grouped as just described.

Conversely, in an abstract set-up where~$\set{A}$ contains
measurements represented by all possible non-degenerate Hermitian
operators, a degenerate Hermitian operator can be regarded as
representing a measurement which is a degenerate form of some
measurement in~$\set{A}$.

\section{Discussion}
\label{sec:discussion}

\subsection{General discussion of the Formulation}

Above, we have formulated a set of background
assumptions~(\emph{partitioning}, \emph{time}, and~\emph{states}),
an abstract experimental set up, and a set of postulates, from
which we have shown that it is possible to derive the
finite-dimensional abstract quantum formalism~(apart from the
explicit form of the temporal evolution operator, which will be
derived in Paper~II).

As described earlier, the background assumptions and the
postulates have been formulated as far as possible so that they
possess the properties of transparency and traceability.  The
background assumptions and a number of the
postulates~(Postulates~3,~3.1,~3.3) are drawn unchanged from the
framework of classical physics, and most of the remaining
postulates are drawn from the framework of classical physics but
modified in light of experimental facts~(Postulate~1.1), or are
based on a classical-quantum correspondence
argument~(Postulates~2.1,~2.4,~3.2,~3.4,~5). Hence, the majority
of the background assumptions and postulates can be traced to
facts or principles that are, or can be, well grounded or
reasonably well grounded in experimental facts or in our
theoretical intuition.

Of the remaining, novel postulates~(Postulates~1.2,~2.2,~2.3,~4),
Postulate~1.2 is a direct generalization of experimental facts, and
Postulate~4 is a reasonable consistency principle. Postulates~2.2
and~2.3 are both transparent in that they can be clearly understood
as assertions about the physical world, and Postulate~2.3 is
traceable to a plausible theoretical principle. Furthermore, since
Postulates~2.2 and~2.3, in conjunction with the above-mentioned
postulates, give rise to the abstract quantum formalism, there is
good reason to believe that they are valid. Nevertheless, these two
postulates, particularly Postulate~2.2, are less well grounded in
our theoretical intuition than the others, and since they play such
a key role in the emergence of the quantum formalism, they shall be
discussed further below.

We mention briefly that it is also possible to understand some of
the postulates using concepts that have not been mentioned thus
far. For example, Postulate~2.1 implies that, when a measurement
is performed on a system, there are degrees of freedom in the
state of a system about which no information is gained.    Hence,
Postulate~2.1 can be regarded as a concrete expression of Bohr's
principle of complementarity. Consequently, it is possible for
different measurements in the measurement set,~\set{A}, to be
inequivalent in that they yield inequivalent information about the
state of the system. If one accordingly regards measurements
in~\set{A} as providing distinct, inequivalent points of view of a
physical system, then two questions arise which do not arise in
classical physics, namely (a)~how should one theoretically
represent these different measurements, and (b)~whether some
measurements yield more information about the state of a system
than other measurements. Postulate~1.2 answers the first question
by asserting that it is possible to represent all measurements
in~\set{A} in terms of any given measurement in~\set{A} and
appropriately chosen interactions in the interaction set,~\set{I}.
Postulate~2.3 answers the second question with the assertion that
none of these points of view are privileged insofar as the
\emph{amount} of information they yield about the system, which
can be regarded as a kind of principle of relativity applied to
the perspectives provided by the different measurements
in~$\set{A}$.

 The derivation itself is noteworthy in
several respects.  First, it gives rise to a mathematical
structure that is neither more nor less general than the
finite-dimensional abstract quantum formalism.  Therefore, any
change to the formalism would require a modification of the
postulates or background assumptions. Consequently, as we shall
illustrate below, the derivation provides an excellent
`laboratory' for investigating proposed modifications of the
quantum formalism.

Second, the derivation yields the conclusion that physical
transformations are represented either by unitary or antiunitary
transformations.  This is a rather remarkable, unanticipated feature
of the derivation since antiunitary transformations are not
generally regarded as an integral part of the abstract quantum
formalism~(as formalized, for instance, by Dirac or von Neumann),
but are instead usually introduced by reference to the theorem of
Wigner~\cite{Wigner-group-theory} mentioned in the Introduction. In
addition, we note that antiunitary transformations have not been
obtained in any of the recent attempts to derive the quantum
formalism in which a significant fraction of the quantum formalism
is obtained~\cite{Caticha98b, Caticha99b, Hardy01a, Hardy01b,
Clifton-Bub-Halvorson03, Grinbaum03, Grinbaum04}.  Furthermore,
since unitary and antiunitary transformations emerge simultaneously
in the above derivation, the derivation suggests that antiunitary
transformations are, in fact, an integral part of the quantum
formalism.

Third, the derivation shows that the use of complex numbers in the
quantum formalism is directly connected with the fact that the set
of possible physical transformations can be represented by the set
of all unitary or antiunitary transformations of a suitably
defined complex vector space. Specifically, the complex form of
the quantum state and the (anti)unitarity of physical
transformations arise simultaneously as a result of imposing
Postulate~3.2 which, in turn, is based on the simple idea that a
change in the overall value of the~$S_i$ in the Hamilton-Jacobi
model has no physically observable consequences.  Hence, the
derivation significantly elucidates the use of complex numbers in
the quantum formalism.

Fourth, it is apparent from the derivation that the concept of
information plays a substantial role in giving rise to the quantum
formalism.  The information gain condition directly leads to
$Q$-space, which introduces square-roots of probability, or
\emph{real} amplitudes and, via Postulate~2.3, leads to a
$2N$-dimensional $Q$-space. Furthermore, in conjunction with
Postulate~2.4, Postulate~2.3 leads to the function~$f(\var_i) =
\pm \cos(a\var_i + b)$.  Hence, the sinusoidal functions into
which the phases in a quantum state enter can be directly traced
to the concept of information.  Finally, the prior over the unit
hypersphere in $Q^{2N}$-space induced by the imposition of
Postulate~2.3 leads, via Postulate~4, to the strong constraint
that physical transformations can only be represented by
orthogonal transformations of the unit hypersphere.

Fifth, the formulation highlights the physical importance of the
notion of a prior over a continuous parameter.  The notion plays a
key role in the derivation, entering through the definition of the
Shannon-Jaynes entropy and through Postulate~2.4.  This is
noteworthy since the notion of prior appears to be
underappreciated, occurring rather infrequently in discussions of
the probabilistic aspects of quantum theory, and not occurring in
most of the aforementioned deductive approaches to quantum
theory~(the approach due to Caticha~\cite{Caticha98b, Caticha99b}
being the only exception).

Sixth, from the perspective provided by the derivation, one can see
rather clearly which assumptions quantum theory shares with
classical physics, which assumptions are modifications of classical
ideas in light of experimental facts, which assumptions are drawn
from classical physics using a correspondence argument, and which
are novel insofar as they have no classical counterparts. In
particular, one can see that the new ideas that need to be
introduced beyond those familiar from classical physics in order to
obtain the quantum formalism all arise from the concepts of
probability, information, or from classical-quantum correspondence
arguments. Since ideas concerning probability and correspondence
played an important role in the historical development of quantum
theory and in its interpretation in the years immediately following
its formulation, the concept of information is the obvious new
addition.

\subsubsection{Discussion of Postulate~2.2.}
\label{sec:discussion-of-postulate-2-2}

Postulate~2.2 introduces the assumption that, when a measurement
is performed on a physical system, there are outcomes~(which we
have labeled~$a$ and~$b$, and~$+$ and~$-$) that are objectively
realized, but go unobserved by the experimenter.

The apparently successful derivation of the quantum formalism lends
support to the plausibility of the assumption that a measurement
generates unobserved outcomes. As mentioned above, the assumption
also has the benefit of transparency.  Nevertheless, it raises two
natural questions, namely~(i) to what physical property or
properties should the outcomes~$a$ and~$b$, and~$+$ and~$-$ be
attributed,~(ii) why are these outcomes not observed in standard
experiments.   A preliminary response to these questions is as
follows.

First, by examining the quantum model of a structureless particle in
the classical limit~(as $m$~tends to macroscopic values), we have
seen that, for a system in an eigenstate of energy, the
variable~$\var_i$ in the quantum model corresponds to~$S_i$ in the
discretized form of the classical Hamilton-Jacobi model. Now,
the~$S_i$ encode the local momenta and total energy of the system.
Hence, if a position measurement is performed and yields the
observed outcome~$i$, then we can associate the outcomes~$a, b$
and~$+, -$ with the local momenta and the total energy of the
system.

More generally, if a measurement~$\mment{A}$ is performed on a
system, it seems reasonable to associate the outcomes~$a, b$ and~$+,
-$ with the property~$A'$, which is complementary to property~$A$,
and with the total energy,~$E$, of the system.  We shall say that
property~$A'$ is complementary to the property~$A$ measured
by~$\mment{A}$ in the sense that exact knowledge of the
properties~$A$ and~$A'$ suffice to determine the classical state of
the system.

Second, the unobservability of the outcomes~$a, b$ and~$+, -$ may be
roughly understood as follows.  We shall see in Paper~II that, for a
system in an eigenstate of energy~$E$, the overall phase,~$\var$, of
its quantum state~(in the complex representation) changes at the
rate~$-E/\hbar$. A measurement which is able to resolve the
outcomes~~$a, b$ and~$+, -$ must therefore have a temporal
resolution~$\Delta t < \hbar/E$.
Now, according to the energy-time uncertainty relation~$\Delta E
\Delta t \geq \hbar/2$~\footnote{%
    We shall regard~$\Delta E \Delta t \geq \hbar/2$ as being a
    consequence of the classical result~$\Delta\omega \Delta t \geq
    1/2$~(relating the uncertainty in the duration and angular
    frequency of a wave) and the photon energy-frequency
    relationship~$E=\hbar \omega$.  However, the validity and meaning
    of the energy-time uncertainty relation, and of the inferences
    that can legitimately drawn from it, have been, and continue to
    be, the subject of debate~(see, for example~\cite{Peres-QT},\S~12.8,
    and~\cite{Oppenheim-PhD}). The
    argument given in the text leading to~$\Delta E \geq E/2$ should,
    accordingly, only be regarded as suggestive insofar as it relies
    on a particular interpretation of the energy-time uncertainty
    relation.},
the energy associated with the interaction used to implement the
measurement has uncertainty~$\Delta E  \geq \frac{1}{2} \hbar/\Delta
t$, so that~$\Delta E \geq E/2$. From~$E=mc^2$, it then follows
that~$\Delta E$ must be of the order of the rest energy of the
system.  A measurement of such energy would therefore probably not
preserve the identity of the system, thereby violating the
assumption that interactions preserve the identity of the
system~(see Sec.~\ref{sec:F-idealised-set-up}). Hence, a measurement
with the requisite temporal resolution cannot be consistently
described within the quantum formalism. Conversely, a measurement
that, with high probability, preserves the identity of the system,
will have insufficient temporal resolution to resolve the
outcomes~$a, b$ and~$+, -$.

\subsubsection{Discussion of Postulate~2.3}
\label{sec:discussion-of-postulate-2-3}

The information gain condition plays a key role in the above
derivation via Postulate~2.3.  In order to obtain a clearer
understanding of the condition, it is helpful to ask whether it
resembles, or is equivalent to, other informational principles, or
has other consequences which coincide with well-known results.
Below, we shall outline two of the consequences which are in
agreement with results that are well-known in probability theory and
statistics, and shall outline the connections to two other
informational principles that have been proposed in the context of
recent informational approaches to quantum theory.

First, we have shown elsewhere~\cite{Goyal05a} that the assumption
that the information gain condition applies to a probabilistic
source is equivalent to Jeffreys' rule~\cite{Jeffreys39}, a
general rule for the assignment of prior probabilities which was
first suggested in the context of probability theory.  This rule
is widely used in some areas~(in econometrics, for example), and
yields priors for parameterized probability distributions~(such as
for the mean and standard deviation of a Gaussian distribution)
that are in agreement with the results of other, independent lines
of argument~(see~\cite{Jaynes68}, for example).

We also note that the metric~$ds^2 = \sum_i dQ_i^2$, introduced in
Sec.~\ref{sec:D1-info-gain}, provides a natural measure of the
distance between probability distributions, and is equivalent, up
to an irrelevant multiplicative constant, to the Fisher
metric,~$ds_F^2 = \sum_i dP_i^2/P_i$, which measures the distance
between the probability distributions~$\vect{P}$ and~$\vect{P} +
\delta \vect{P}$.

Second, we note that the Fisher metric was obtained
in~\cite{Wootters-statistical-distance} as a natural measure of
the distance between probability distributions, where it was
connected with the Hilbert space distance between pure states. The
Fisher metric also gives rise to the so-called Fisher information
of a continuous probability distribution, which is central to the
Fisher information approach to understanding quantum
theory~\cite{Frieden-Schroedinger-derivation,
Reginatto-Schroedinger-derivation}.

Finally, we note that, if the information gain condition applies
to a probabilistic source with some probability
n-tuple,~$\vect{P}$, it follows that, in~$n$ interrogations of the
source, the amount of Shannon-Jaynes information provided by the
data about~$\vect{P}$ is an increasing function of~$n$ in the
limit as~$n \rightarrow \infty$.  This condition, which we shall
call the \emph{condition of information increase}, accords with
the rather simple and intuitively plausible idea that, as one
gathers more data from a probabilistic source, one's information
about~$\vect{P}$ strictly increases.  This condition was first
proposed, in a slightly different form, in~\cite{Summhammer99},
where it forms the basis for an attempt to derive a part of the
quantum formalism.

Hence, it appears that the information gain condition has a number
of interesting and important connections to results in probability
theory and to principles in various informational approaches to
quantum theory.

\subsection{Some Implications of the Deduction}

\subsubsection{Information in Quantum Theory}

One of the major objectives of the programme of deriving quantum
theory using the concept of information is to determine whether the
concept of information is indispensable to our understanding of the
quantum formalism, and, if so, to illuminate the precise
relationship between the concept of information and the quantum
formalism.

On the first issue, although many recent approaches to derive the
quantum formalism involve the concept of information, the
conclusion that information is indispensable to our understanding
of the quantum formalism cannot be drawn, either because the
approaches are unable to obtain the quantum formalism~(even though
they are able to derive specific results, such as Malus' law), or
because, in those approaches that are able to obtain a significant
fraction of the quantum formalism,  the abstract nature of some of
the assumptions that are employed obscures the role played by
information in determining the formalism.  Indeed, further doubt
on the need for information is cast by other recent approaches,
most notably due to Hardy~\cite{Hardy01a, Hardy01b}, that are
successful in deriving a significant fraction of the quantum
formalism without invoking the concept of information in any way.

On the second issue, it is remarkable that the manner in which the
concept of information is formalized differs considerably amongst
the various informational approaches. Consequently, as we shall
elaborate upon below, the question of precisely \emph{how} one
should formalize the concept of information in the quantum setting
has received a wide range of often incompatible answers. However,
it is difficult to evaluate the relative merits of these answers,
for the same reasons just given above, namely either because the
approaches are too incomplete or because they use abstract
assumptions that obscure the role played by information.

The formulation presented here provides significant new insight
into both of these issues.  First, the formulation rests on
assumptions that are transparent and that are, to a large extent,
traceable to familiar or well-established experimental facts or
theoretical ideas.  For example,  abstract assumptions that
directly introduce complex numbers are avoided. As a result, the
role played by information in the derivation can be clearly seen,
and its role is sufficiently widespread that it seems very likely
that the concept of information could indeed have a fundamental
role to play in our understanding of the origin of the quantum
formalism.

In order to discuss the second issue, it is convenient to classify
the above-mentioned differences in the formalization of the
concept of information with respect to (a)~what the information is
about, (b)~whether or not information is quantified in some way,
(c)~which information measure is chosen, and (d)~when the
Shannon-Jaynes measure is used, whether there is a naturally
preferred prior, and, if so, what is the form of the prior.

In particular, with respect to~(a), in~\cite{Wootters80},
information gain is, as in our approach, regarded as the gain of
information about the state of the system due to the receipt of
data obtained through performing a measurement on the system.  In
contrast, in~\cite{Brukner02b}, information gain is taken to be
the removal of the uncertainty of the experimenter about the
outcome of a measurement as a result of the measurement being
performed. In respect to~(b), one finds that, for example,
in~\cite{Popescu-Rohrlich97, Clifton-Bub-Halvorson03}, information
is not subject to quantification, whereas
in~\cite{Brukner02b,Wootters80}, a particular quantification
measure is employed.

With respect to~(c), the Shannon-Jaynes entropy is used
in~\cite{Wootters80}, whereas~\cite{Brukner02b} employs a measure
that differs from the Shannon entropy, it being argued that the
Shannon entropy is inapplicable in the quantum
setting~\cite{Brukner02a}.  Finally, with respect to~(d), some
authors~\cite{Fuchs02} appear to hold the view that there is no
natural basis for determining a prior for the Shannon-Jaynes
entropy, while, in the field of probability theory, authors who have
sought plausible general principles for the assignment of priors
have obtained different priors over probability n-tuples~(for
example, see~\cite{Jeffreys39, Jaynes68}) on the basis of their
arguments.

The approach described here supports the view that information is
primarily to be regarded as information gained about the state of
a system by an experimenter as a result of performing measurements
on the system. In addition, the approach demonstrates the
importance of information quantification, and provides significant
support for the view that the Shannon-Jaynes entropy is the
appropriate information measure in a quantum setting.

Finally, we have shown that, for an experimenter who receives a
system prepared in a pure but unknown state, it is possible to
formalism an intuitively plausible principle~(Postulate~2.3) which
determines the prior for the probabilistic source that models a
measurement performed on the system by the experimenter. As
described in Sec.~\ref{sec:F-idealised-set-up}, one can see that the
experimenter's state of knowledge in this case is not arbitrarily
chosen, but precisely reflects the knowledge that a system has been
prepared in such a way that its pre-preparation history is
irrelevant insofar as the outcomes of subsequent measurements in the
set-up are concerned~(a preparation which is analogous to an
idealized complete preparation in classical physics) and therefore
has fundamental physical significance.

\subsubsection{Interpretation and Modification of Quantum Theory}

The deductive formulation has several implications for some issues
of concern in the interpretation of quantum theory, and for some
of the proposed modifications of quantum theory.  We shall briefly
outline one example.

\medskip
\paragraph{Modification of the Quantum Formalism.}  Since the
development of the quantum formalism, there has been some
uncertainty as to whether the formalism is the most general
formalism for the description of quantum phenomena. Various
possibilities have been suggested for the generalization of the
formalism which, from a purely mathematical point of view, seem to
be plausible, and which may have interesting physical consequences.
For example, the possibility of non-unitary temporal evolution has
been considered by several authors~\cite{Weinberg89a, Weinberg89b,
Herbert82}.

In some cases, it is possible to devise experimental tests to rule
out certain types of modification on physical grounds.  However,
it is not always possible to devise such tests or to implement
them.  The deductive formulation described here provides another
way in which the physical plausibility of a proposed modification
may be assessed.

The deductive formulation shows that a set of postulates implies the
existing quantum formalism. Hence, if any proposed modification of
the formalism is to be valid, one or more of these postulates must
be changed in some way. By tracing the dependency of the features of
the quantum formalism that are at issue to specific postulates, and
assessing the consequences of modifying one or more of these
postulates, one can potentially use the deductive formulation to
obtain another indication as to whether a proposed modification is
physically plausible. Furthermore, the formulation has the potential
to allow one to explicitly work out the effect that specific changes
to particular postulates would have upon the quantum formalism.

For example, for the purpose of illustrating how the deductive
formulation can help guide modifications to quantum formalism,
suppose that one wishes to modify the quantum formalism so as to
allow continuous transformations to be represented by non-unitary
transformations.  Now, in the deductive formulation, unitarity
depends most directly upon Postulate~3.2~(\emph{Invariance}), and
additionally depends upon several supporting postulates which are
based on classical physics, on probabilistic ideas, or on novel
assumptions. The proposed modification implies that one or more of
these postulates needs to be modified.

Amongst the supporting postulates, all but Postulate~2.2 have a
reasonably high degree of certainty. However, it does not appear to
be possible to modify Postulate~2.2 in any plausible manner so as to
give rise to non-unitary transformations.  The most likely candidate
for modification therefore appears to be Postulates~3.2.

Consider the extreme case where the constraint imposed by
Postulate~3.2 is entirely removed. Then, the set of possible
transformations consists of the set of orthogonal transformations
of the unit hypersphere in~$Q^{2N}$. When expressed in complex
form, this set of transformations contains transformations that
are neither unitary nor antiunitary.  Thus, a simple modification
of the postulates readily yields a set of non-unitary
transformations which can then be subjected to further examination
to assess their physical significance and plausibility.

\section{Conclusion}

In this paper, we have shown that majority of the finite-dimensional
abstract quantum formalism can be derived from a set of physically
comprehensible assumptions. The derivation illuminates the physical
origin of the quantum formalism and the role played by information
in quantum theory, makes clearer the commonalities and differences
in the assumptions underlying quantum physics and classical physics,
and potentially has significant implications for the interpretation
and proposed modifications of quantum theory.

\begin{acknowledgments}

I am indebted to Steve Gull and Mike Payne for their constant
support and encouragement, and to Tetsuo Amaya for extensive
critical comments and suggestions.  I am also indebted to Ariel
Caticha, Matthew Donald, Chris Fuchs, Suguru Furuta, Lucien Hardy,
Kevin Knuth, and John Skilling for discussions and invaluable
comments.
\end{acknowledgments}

\newpage
\appendix

\section{Implementation of the Information Gain Condition}

In this appendix, we shall more formally implement the information
gain condition~(Sec.~\ref{sec:D1-info-gain}) in the general case
of an $M$-outcome probabilistic source.

First, we parameterize the n-tuple~$\vect{P}$ by
the~$(M-1)$--dimensional parameter n-tuple~$\veclam = (\lamvec)$, so
that~$\vect{P} = \vect{P}(\veclam)$, where the parametrization is
invertible and differentiable, and then set the prior
probability,~$\Pr(\veclam | \text{I})$, equal to a constant.

Next, we determine~$\Pr(\veclam | \vect{f},n,\text{I})$. From
Bayes' theorem, the posterior probability is given by
\begin{equation} \label{eqn:posterior-over-lambda}
\begin{split}
\Pr(\veclam|\vect{f},n,\text{I}) &=
                        \frac{ \Pr(\vect{f}|\veclam,
                        n, I) \Pr(\veclam | n,\text{I})
                        }{\idotsint  \Pr(\vect{f}|\veclam,
                        n, I) \Pr(\veclam | n,\text{I}) \,d\lambda_1 \dots
                        \,d\lambda_{M-1}}                           \\
                        &=
                        \frac{ \Pr(\vect{f}|\veclam,
                        n, \text{I})}{\idotsint  \Pr(\vect{f}|\veclam,
                        n, \text{I})  \,d\lambda_1 \dots \,d\lambda_{M-1}}.  \\
\end{split}
\end{equation}
Here, we have used the fact that~$\Pr(\veclam |n, \text{I}) =
\Pr(\veclam |\text{I})$. This follows from an application of
Bayes' theorem, $\Pr(\veclam|n,\text{I}) \Pr(n|\text{I}) =
\Pr(n|\veclam, \text{I}) \Pr(\veclam| \text{I})$, and the fact
that~$n$ is chosen freely by the experimenter and therefore cannot
depend upon~$\veclam$. Hence, the posterior probability is
proportional to the likelihood,~$\Pr(\vect{f} |\veclam, n,
\text{I})$.

When~$n$ is large, using Stirling's approximation,~$n! = n^n (2\pi
n)^{1/2}e^{-n} + \text{O}(1/n)$, the
likelihood~(Eq.~\eqref{eqn:likelihood}) becomes
\begin{equation} \label{eqn:likelihood-approximation}
\begin{split}
\Pr(\vect{f} |\veclam, n, \text{I})
                    &= \frac{(2\pi n)^{1/2}} {(2\pi n)^{M/2}}
                    \frac{1}{\sqrt{f_1 f_2 \dots f_M}}
                    \prod_i
                    \left(\frac{P_i(\veclam)}{f_i}\right)^{nf_i}\\
                    &= \frac{(2\pi n)^{1/2}} {(2\pi n)^{M/2}}
                    \frac{1}{\sqrt{f_1 f_2 \dots f_M}} \\
                    & \quad\quad\quad\quad\quad \times \exp \left( -n \sum_i f_i \ln \frac{f_i}{P_i(\veclam)} \right). \\
\end{split}
\end{equation}
In the limit of large~$n$, the
posterior,~$\Pr(\veclam|\vect{f},n,\text{I})$ is sharply peaked
about~$\veclam^{(0)}$, defined by~$\vect{f} =
\vect{P}(\veclam^{(0)})$.  To find the form of the posterior
about~$\veclam^{(0)}$, we expand the likelihood
about~$\veclam^{(0)}$.  We write
\begin{equation}
P_i(\veclam) =
            P_i(\veclam^{(0)}) +
            \sum_{l=1}^{M-1} \frac{\partial P_i}{\partial \lambda_l}
            \Bigg\vert_{\veclam^{(0)}}
            (\lambda_l - \lambda_l^{(0)}) + \dots,
\end{equation}
and note that
\begin{widetext}
\begin{equation}
\begin{split}
    \sum_i f_i \ln \left(\frac{P_i(\veclam)}{f_i} \right)
    &=
    \sum_i f_i \ln\left( 1+ \frac{1}{f_i} \sum_l \frac{\partial
    P_i}{d\lambda_l} (\lambda_l - \lambda_l^{(0)}) +
    \frac{1}{2f_i} \sum_{l,l'} \frac{\partial^2 P_i}{d\lambda_l
    d\lambda_{l'}} (\lambda_l - \lambda_l^{(0)})
    (\lambda_{l'} - \lambda_{l'}^{(0)}) + \dots \right) \\
    &=
    \sum_i f_i \left( \frac{1}{f_i} \sum_l \frac{\partial
    P_i}{d\lambda_l} (\lambda_l - \lambda_l^{(0)}) +
    \frac{1}{2f_i} \sum_{l,l'} \frac{\partial^2 P_i}{d\lambda_l
    d\lambda_{l'}} (\lambda_l - \lambda_l^{(0)})
    (\lambda_{l'} - \lambda_{l'}^{(0)}) + \dots \right) \\
    & -\sum_i \frac{f_i}{2}
    \left( \frac{1}{f_i} \sum_l \frac{\partial
    P_i}{d\lambda_l} (\lambda_l - \lambda_l^{(0)}) +
    \frac{1}{2f_i} \sum_{l,l'} \frac{\partial^2 P_i}{d\lambda_l
    d\lambda_{l'}} (\lambda_l - \lambda_l^{(0)})
    (\lambda_{l'} - \lambda_{l'}^{(0)}) + \dots \right)^2
    + \dots \\
    &= \left[ P_i(\veclam) - P_i(\veclam^{(0)})\right] \\
      &\quad -\frac{1}{2} \sum_l \sum_{l'} \sum_i
     \frac{1}{f_i} \frac{\partial P_i}{d\lambda_l}
                    \frac{\partial P_i}{d\lambda_{l'}}
     (\lambda_l - \lambda_l^{(0)}) (\lambda_{l'} - \lambda_{l'}^{(0)})
    + O\left((\lambda_l - \lambda_l^{(0)})^3\right) \\
    &=  -\frac{1}{2} \sum_l \sum_{l'} \sum_i
     \frac{1}{f_i} \frac{\partial P_i}{d\lambda_l}
                    \frac{\partial P_i}{d\lambda_{l'}}
     (\lambda_l - \lambda_l^{(0)}) (\lambda_{l'} - \lambda_{l'}^{(0)})
    + O\left((\lambda_l - \lambda_l^{(0)})^3\right),
\end{split}
\end{equation}
\end{widetext}
where the~$\ln$~term has been expanded out and we have used the
fact that~$\sum_i P_i = 1$.  Retaining only the leading order
terms in the~$\lambda_l$, the likelihood becomes
\begin{equation}
\begin{split}
\Pr(\vect{f} |\veclam, n, \text{I})
                    &=
                    \frac{(2\pi n)^{1/2}} {(2\pi n)^{M/2}}
                    \frac{1}{\sqrt{f_1 f_2 \dots f_M}} \\
                    & \prod_{l=1}^{M-1} \prod_{l'=1}^{M-1}
                    \exp \left(
                    - \frac{ (\lambda_l - \lambda_l^{(0)} ) (\lambda_{l'} - \lambda_{l'}^{(0)}
                    )}{2 \sigma_{ll'}^2 }
                    \right),
\end{split}
\end{equation}
where
\begin{equation} \label{eqn:sigmajk}
\frac{1}{\sigma_{ll'}^2} = n \sum_{i=1}^M
                                    \frac{1}{P_i(\veclam^{(0)})} \frac{\partial P_i}{\partial
                                    \lambda_l}\Bigg\vert_{\veclam^{(0)}}
                                    \frac{\partial P_i}{\partial \lambda_{l'}}
                                    \Bigg\vert_{\veclam^{(0)}}.
\end{equation}
The posterior can then be obtained from
Eq.~\eqref{eqn:posterior-over-lambda}.  For example, in the case
where~$M=2$,
\begin{equation} \label{eqn:posterior-over-lambda-M-is-2}
\begin{split}
\Pr(\lambda_1|\vect{f},n,\text{I}) &=
                        \frac{\Pr(\vect{f}|\lambda_1,
                        n, I)}{\int  \Pr(\vect{f}|\lambda_1,
                        n, I)  \,d\lambda_1}  \\
                        &= \frac{1}{\sigma_{11}\sqrt{2\pi}}
                        \exp\left(-\frac{(\lambda_1 -
                        \lambda_1^{(0)})^2}{2\sigma_{11}^2}\right),
\end{split}
\end{equation}
and, more generally,
\begin{equation} \label{eqn:lambda-posterior}
\begin{split}
\Pr(\veclam |\vect{f}, n, \text{I}) &=
                \frac{(\det B)^{1/2}}{(2\pi)^{(M-1)/2}} \\
                &\times\prod_{l=1}^{M-1} \prod_{l'=1}^{M-1}
                \exp \left(
                    - \frac{ (\lambda_l - \lambda_l^{(0)} ) (\lambda_{l'} - \lambda_{l'}^{(0)}
                    )}{2 \sigma_{ll'}^2 }
               \right),
\end{split}
\end{equation}
where~$B_{ll'} = 1/\sigma_{ll'}^2$.

Now, consider an $M$-dimensional real Euclidean space,~$Q^M$, with
axes~$Q_1, Q_2, \dots, Q_M$.  If we define the vector~$\vect{Q} =
(Q_1, Q_2, \dots, Q_M)$ such that~$Q_i = \sqrt{P_i}$~($0\leq Q_i
\leq 1$), then every~$\vect{Q}$ that represents a probability
n-tuple lies on the positive orthant,~$\orthant$, of the unit
hypersphere,~$\hypersphere$. Eq.~\eqref{eqn:sigmajk} can be then
rewritten as
\begin{equation}
\begin{split}
        \frac{1}{\sigma_{ll'}^2}     &= 4n \sum_{i=1}^M
                                    \frac{\partial Q_i}{\partial \lambda_l} \Bigg\vert_{\veclam^{(0)}}
                                    \frac{\partial Q_i}{\partial \lambda_{l'}} \Bigg\vert_{\veclam^{(0)}}.    \\
\end{split}
\end{equation}
For example, in the case where~$M=2$,
\begin{equation}
\begin{split}
        \frac{1}{\sigma_{11}^2}     &= 4n
                                    \left[
                                    {\left(\frac{dQ_1}{d\lambda_1} \right)
                                    }^2 \Bigg\vert_{\lambda_1^{(0)}}
                                    +
                                     {\left( \frac{dQ_2}{d\lambda_1} \right)
                                    }^2 \Bigg\vert_{\lambda_1^{(0)}} \right]   \\
                                    &= 4n
                                   {\left(
                                      \frac{ds}{d\lambda_1}
                                    \right)}
                                   ^2 \Bigg\vert_{\lambda_1^{(0)}}, \\
\end{split}
\end{equation}
where~$ds^2 = dQ_1^2 + dQ_2^2$ is the metric in $Q^2$. The
posterior,~$\Pr(\lambda_1 | \vect{f},n,I)$, is therefore a
Gaussian with standard deviation,
\begin{equation} \label{eqn:lambda-sigma}
\sigma = \frac{1}{2\sqrt{n}}
{\left(\frac{ds}{d\lambda_1}\right)}^{-1}
\Bigg\vert_{\lambda_1^{(0)}},
\end{equation}
where~$s$ is the distance along the positive quadrant of the unit
circle.  Since~$\Pr(\lambda_1 | \text{I})$ is constant,
\begin{equation}
\begin{split}
    \Delta K   &= \frac{1}{2} \ln \left(\frac{2n}{ \pi e} \right)
                    +  \ln      {\left|
                            \frac{ds}{d\lambda_1}
                                \right|}
                                \Bigg\vert_{\lambda_1^{(0)}}
                    - \ln \left[ \Pr(\lambda_1|\text{I}) \right] \\
                &= \frac{1}{2} \ln \left(\frac{2n}{ \pi e}\right)
                    - \ln \left[ \Pr \right(s(\lambda_1^{(0)})|\text{I} \left) \right]
\end{split}
\end{equation}
where the relation~$\Pr(\lambda_1|\text{I})|d\lambda_1| =
\Pr(s|\text{I})|ds|$ has been used to arrive at the second line.
Independence of~$\Delta K$ from~$f_1$ can be ensured if and only
if~$\Pr(s|\text{I})$ at~$\lambda_1^{(0)}$ is a constant
on~$\orthant$, where the constant is non-zero in order to ensure
that the parametrization of~$\vect{P}$ is invertible.  In this case,
\begin{equation} \label{eqn:Delta-H-final}
\Delta K = \frac{1}{2} \ln \left(\frac{2n}{\pi e} \right) +
\text{const.}
\end{equation}

Since we assumed at the outset that~$\Pr(\lambda_1|\text{I})$ is a
constant on~$\orthant$, it follows from the relation
\begin{equation} \label{eqn:lambda-s-reln}
\Pr(\lambda_1|\text{I})|d\lambda_1| = \Pr(s|\text{I})|ds|
\end{equation}
that~$s(\lambda_1) = a \lambda_1 + b$, where~$a, b$ are arbitrary
real constants. From Eq.~\eqref{eqn:lambda-sigma}, it then follows
that~$\sigma = 1/2a\sqrt{n}$, and, from
Eq.~\eqref{eqn:lambda-s-reln}, it then follows that the posterior
over the positive quadrant of the unit circle is a Gaussian whose
standard deviation is~$1/2\sqrt{n}$, which is independent
of~$\vect{Q}$.

The treatment for general~$M$ runs parallel to the above. Suppose
that the~$\lambda_l$ are chosen such that infinitesimal changes in
the~$\lambda_l$ generate orthogonal displacements in~$Q^N$--space.
This can be done by using hyperspherical
co-ordinates,~$(r,\theta_1, \theta_2, \dots, \theta_{M-1})$,
with~$r=1$ and, for~$l=1, \dots, M-1$, with~$\theta_l$ being a
function of~$\lambda_l$ only.  In that case, one finds that
\begin{equation} \label{eqn:sigmall'final}
\sigma_{ll'} =
     \frac{1}{2\sqrt{n}} {\left( \frac{\partial s}{\partial \lambda_l} \right)^{-1}}\bigg\vert_{\veclam^{(0)}}
     \,\,\delta_{l,l'}.
\end{equation}
Consequently, the posterior
probability~(Eq.~\eqref{eqn:lambda-posterior}) reduces to a
product of Gaussian functions,
\begin{equation}
\Pr(\veclam |\vect{f}, n, \text{I}) =
                \prod_{l=1}^{M-1}
                \frac{1}{\sigma_{ll} \sqrt{2\pi}}
                \exp \left(
                    - \frac{ (\lambda_l - \lambda_l^{(0)} )^2
                    }{2 \sigma_{ll}^2 }
                    \right),
\end{equation}
and the information gain becomes
\begin{equation} \label{eqn:delta-H-general}
\begin{split}
   \Delta K     &= -\sum_{l=1}^{M-1} \ln \bigl( \sigma_{ll} \sqrt{2 \pi e} \bigr)  \\
                &= \frac{N-1}{2} \ln \left( \frac{2n}{\pi e}
                \right)
                    + \sum_{l=1}^{M-1} \ln \frac{\partial s}{\partial \lambda_l}
                            \Bigg\vert_{\veclam^{(0)}} \\
                    &\quad\quad\quad\quad
                         - \ln \left[ \Pr(\lambda_1, \lambda_2, \dots,
                                    \lambda_{M-1}|\text{I}) \right] \\
                &= \frac{M-1}{2} \ln \left(\frac{2n}{\pi e}
                \right)
                     - \ln \left[ \Pr(s_1, s_2, \dots, s_{M-1}|\text{I}) \right],
\end{split}
\end{equation}
where~$ds^2 = dQ_1^2 + dQ_2^2 + \dots + dQ_M^2$ and where~$ds_l =
(\partial s / \partial \lambda_l)|_{\veclam^{(0)}} d\lambda_l$.

Since the $\lambda_l$ are independent variables, independence
of~$\Delta K$ from the~$\lambda_l$ can be ensured if and only if
the prior~$\Pr(s_1, s_2, \dots, s_{M-1}|\text{I})$ is a constant
on~$\orthant$ independent of the~$\lambda_l$, in which case
\begin{equation}
\Delta K = \frac{M-1}{2}\ln\left(\frac{2n}{\pi e}\right) +
\text{const.}
\end{equation}
Therefore, any area element, $dA = \prod_{l=1}^{M-1}
d\,s_l$, on~$\orthant$ is weighted proportionally to its area
independent of its location on the unit hypersphere. Hence, the
information gain condition is equivalent to the condition that the
prior over~$\orthant$ is uniform.

From the constancy of~$\Pr(s_1, s_2, \dots, s_{M-1}|\text{I})$
derived above, it follows that~$\Pr(s_1|\text{I}),
\Pr(s_2|\text{I}), \dots, \Pr(s_{M-1}|\text{I})$ are all constant.
Similarly, from the constancy of~$\Pr(\lambda_1, \lambda_2, \dots,
\lambda_{M-1}|\text{I})$, which we assumed at the outset, follows
the constancy of the~$\Pr(\lambda_l|\text{I})$. From the
relations~$\Pr(\lambda_l|\text{I})d\lambda_l =
\Pr(s_l|\text{I})ds_l$~($l=1, 2, \dots, M-1$), it then follows
that
    \begin{equation} \label{eqn:reln-of-s-to-lambda}
    s_l = a_l \lambda_l + b_l,
    \end{equation}
where the~$a_l$ and~$b_l$ are arbitrary constants.   From
Eq.~\eqref{eqn:sigmall'final}, we obtain that
\begin{equation}
\sigma_{ll'} = \frac{1}{2a_l\sqrt{n}} \,\,\delta_{l,l'},
\end{equation}
which, using Eq.~\eqref{eqn:reln-of-s-to-lambda}, implies that the
posterior over~$\orthant$ is a symmetric Gaussian function whose
standard deviation is~$1/2\sqrt{n}$, independent of~$\vect{Q}$.

\end{document}